\documentclass[11pt,a4paper,showpacs,showkeys,,superscriptaddress]{article}

\usepackage{epsfig}
\usepackage{amsmath,amssymb}
\usepackage{graphicx}
\usepackage{xcolor}
\usepackage{subfigure}
\usepackage{amstext}
\usepackage{mathrsfs}

\usepackage{latexsym}
\usepackage{hyperref}
\hypersetup{colorlinks,%
  citecolor=blue,%
  linkcolor=red,%
  pdftex}

\begin{document}

\baselineskip 6mm

\newcommand{\nc}{\newcommand}
\newcommand{\rnc}{\renewcommand}



\newcommand{\tcb}{\textcolor{blue}}
\newcommand{\tcr}{\textcolor{red}}
\newcommand{\tcg}{\textcolor{green}}


\def\beq{\begin{equation}}
\def\eeq{\end{equation}}
\def\ba{\begin{array}}
\def\ea{\end{array}}
\def\bea{\begin{eqnarray}}
\def\eea{\end{eqnarray}}
\def\nn{\nonumber}


\def\CMP{Commun. Math. Phys.~}
\def\JHEP{JHEP~}
\def\Pre{Preprint}
\def\PRL{Phys. Rev. Lett.~}
\def\PR {Phys. Rev.~}
\def\CQG {Class. Quant. Grav.~}
\def\PL {Phys. Lett.~}
\def\NP {Nucl. Phys.~}

\def\G{\Gamma}

\def\S{{\bf S}}
\def\C{{\bf C}}
\def\Z{{\bf Z}}
\def\R{{\bf R}}
\def\N{{\bf N}}
\def\M{{\bf M}}
\def\P{{\bf P}}
\def\bm{{\bf m}}
\def\bn{{\bf n}}

\def\CA{{\cal A}}
\def\CB{{\cal B}}
\def\CC{{\cal C}}
\def\CD{{\cal D}}
\def\CE{{\cal E}}
\def\CF{{\cal F}}
\def\CH{{\cal H}}
\def\CM{{\cal M}}
\def\CG{{\cal G}}
\def\CI{{\cal I}}
\def\CJ{{\cal J}}
\def\CL{{\cal L}}
\def\CK{{\cal K}}
\def\CN{{\cal N}}
\def\CO{{\cal O}}
\def\CP{{\cal P}}
\def\CQ{{\cal Q}}
\def\CR{{\cal R}}
\def\CS{{\cal S}}
\def\CT{{\cal T}}
\def\CU{{\cal U}}
\def\CV{{\cal V}}
\def\CW{{\cal W}}
\def\CX{{\cal X}}
\def\CY{{\cal Y}}
\def\CZ{{\cal Z}}

\def\We{{W_{\mbox{eff}}}}


\newcommand{\Lie}{\pounds}

\newcommand{\p}{\partial}
\newcommand{\bp}{\bar{\partial}}

\newcommand{\half}{\frac{1}{2}}

\newcommand{\bfalpha}{{\mbox{\boldmath $\alpha$}}}
\newcommand{\bfbeta}{{\mbox{\boldmath $\beta$}}}
\newcommand{\bfgamma}{{\mbox{\boldmath $\gamma$}}}
\newcommand{\bfmu}{{\mbox{\boldmath $\mu$}}}
\newcommand{\bfpi}{{\mbox{\boldmath $\pi$}}}
\newcommand{\bfvarpi}{{\mbox{\boldmath $\varpi$}}}
\newcommand{\bftau}{{\mbox{\boldmath $\tau$}}}
\newcommand{\bfeta}{{\mbox{\boldmath $\eta$}}}
\newcommand{\bfxi}{{\mbox{\boldmath $\xi$}}}
\newcommand{\bfkappa}{{\mbox{\boldmath $\kappa$}}}
\newcommand{\bfepsilon}{{\mbox{\boldmath $\epsilon$}}}
\newcommand{\bfTheta}{{\mbox{\boldmath $\Theta$}}}

\newcommand{\bz}{{\bar{z}}}

\newcommand{\dalpha}{\dot{\alpha}}
\newcommand{\dbeta}{\dot{\beta}}
\newcommand{\blambda}{\bar{\lambda}}
\newcommand{\btheta}{{\bar{\theta}}}
\newcommand{\bsigma}{{{\bar{\sigma}}}}
\newcommand{\bepsilon}{{\bar{\epsilon}}}
\newcommand{\bpsi}{{\bar{\psi}}}


\def\ct{\cite}
\def\la{\label}
\def\eq#1{(\ref{#1})}


\def\a{\alpha}
\def\b{\beta}
\def\g{\gamma}
\def\G{\Gamma}
\def\d{\delta}
\def\D{\Delta}
\def\ep{\epsilon}
\def\e{\eta}
\def\ph{\phi}
\def\Ph{\Phi}
\def\ps{\psi}
\def\Ps{\Psi}
\def\k{\kappa}
\def\l{\lambda}
\def\L{\Lambda}
\def\m{\mu}
\def\n{\nu}
\def\th{\theta}
\def\Th{\Theta}
\def\r{\rho}
\def\s{\sigma}
\def\S{\Sigma}
\def\ta{\tau}
\def\o{\omega}
\def\O{\Omega}
\def\pr{\prime}
\def\f{\varphi}


\def\half{\frac{1}{2}}

\def\goto{\rightarrow}

\def\na{\nabla}
\def\grad{\nabla}
\def\curl{\nabla\times}
\def\div{\nabla\cdot}
\def\pa{\partial}

\def\bra{\left\langle}
\def\ket{\right\rangle}
\def\lb{\left[}
\def\lc{\left\{}
\def\ls{\left(}
\def\lp{\left.}
\def\rp{\right.}
\def\rb{\right]}
\def\rc{\right\}}
\def\rs{\right)}
\def\cl{\mathcal{l}}

\def\vac#1{\mid #1 \rangle}

\def\td#1{\tilde{#1}}
\def\check{ \maltese {\bf Check!}}


\def\Tr{{\rm Tr}\,}
\def\det{{\rm det}\,}


\def\bc#1{\nnindent {\bf $\bullet$ #1} \\ }
\def\ch {$<Check!>$ }
\def\ss {\vspace{1.5cm}}

\begin{titlepage}
%
%
%
%
%
%
%
%
\begin{center}
{\Large \bf Canonical energy and  hairy AdS black holes}
%
\vskip 1. cm
  {  Seungjoon Hyun\footnote{e-mail : sjhyun@yonsei.ac.kr}, Sang-A Park\footnote{e-mail : sangapark@yonsei.ac.kr},
  Sang-Heon Yi\footnote{e-mail : shyi@yonsei.ac.kr} 
 }
\vskip 0.5cm
{\it Department of Physics, College of Science, Yonsei University, Seoul 120-749, Korea}\\
\end{center}
\thispagestyle{empty}
\vskip1.5cm
%
%
\centerline{\bf ABSTRACT} \vskip 4mm
 \vspace{1cm} 
\noindent  We propose the modified version of the canonical energy which was introduced  originally by Hollands and Wald. Our construction depends only on the Euler-Lagrange expression of the system and thus is independent of the ambiguity in the Lagrangian. After some comments on our construction,  we briefly mention on the relevance of our construction to the boundary information metric in the context of the AdS/CFT correspondence. We also study the stability of three-dimensional hairy extremal black holes by using our construction.

\vspace{2cm}
%
%
\end{titlepage}
\renewcommand{\thefootnote}{\arabic{footnote}}
\setcounter{footnote}{0}
%
%
%
%

\section{Introduction}
Black hole stability has been one of the important persisting issues in black hole physics, whose study has realistic implications, for instance, that the observed black holes would be stable ones if they are found experimentally. This also has interesting applications in the context of the AdS/CFT correspondence   since the stability or unitarity of the finite temperature field theory system would be dual to the stability of black holes or branes according to the AdS/CFT dictionary. It has been well known that there are at least two kinds of stability concepts in black holes, one of which is known as the dynamical stability and the other is the thermodynamic one. At the linearized level, the dynamical stability is determined by mode analysis for perturbing black hole solutions, while the thermodynamics stability is concerned about the stability of black holes relative to other thermodynamic states in an appropriate ensemble. Interestingly, it has been conceived that two kinds of stability have different natures and so do not coincide in general. However, there was a conjecture by Gubser and Mitra that the thermodynamics instability implies the dynamical one at least for black branes~\cite{Gubser:2000ec,Gubser:2000mm}. Since this conjecture relates two different kinds of analysis on black holes, it may indicate a possibility of another approach to the dynamical stability  different from the standard mode analysis.

Recently, another method for the linear dynamical stability of black holes was developed  by Hollands and Wald(HW)\cite{Hollands:2012sf}, which uses the machinery  in the covariant phase space  through the second variation of the covariant symplectic form, named the {\it canonical energy}. By using the canonical energy method, HW have proved the Gubser-Mitra conjecture and showed the consistency of their method with another important criterion for the black hole stability known as the local Penrose inequality~\cite{Figueras:2011he}. This canonical energy method has been applied successfully to the extremal black holes and the asymptotic AdS space~\cite{Hollands:2014lra,Green:2015kur,Keir:2013jga}. More recently, the HW canonical energy in AdS space is conjectured to be dual to the Fisher information metric on the dual quantum system~\cite{Lashkari:2015hha}. This conjecture is based on several interesting properties of the HW canonical energy and is checked explicitly in concrete examples. 

Though the HW canonical energy method has a great advantage in some aspects over the standard mode analysis, it raises the following question. Basically, the HW canonical energy method is not based on the equations of motion(EOM) but on the Lagrangian, while  other methods for the  stability criterion utilize (linearized) EOM or the solutions, themselves. Therefore,  it seems to be better if we could construct the canonical energy or the modified canonical energy by using EOM or the Euler-Lagrange expression, not the Lagrangian. This construction may be relevant especially when only the EOM are known, for example, as in the type IIB supergravity case. In this paper, we attempt to construct the modified version of the HW canonical energy by using the Euler-Lagrange expression, only. Through this construction, it is realized that the canonical energy may have some freedom in its definition, which may be relevant in its interpretation as the dual to the Fisher information metric. 

The paper is organized as follows. 
In  section  two, we review on the quasi-local formalism for charges developed in~\cite{Kim:2013zha,Kim:2013cor,Hyun:2014kfa,Hyun:2014sha,Hyun:2014nma}, which may be regarded as the EOM alternative to the covariant phase space approach. Then, we introduce  a modified canonical energy based on the quasi-local Abbott-Deser-Tekin(ADT) formalism and see its relation to the canonical energy introduced by HW~\cite{Hollands:2012sf} in section three. In pure Einstein gravity, we show that our modified canonical energy leads to the bulk expression only, in contrast to the HW canonical energy. In fact, it turns out that one can add the boundary terms without destroying the properties for the canonical energy and so this seems to indicate that there is freedom in the definition of the canonical energy. In section four, we study the stability of  extremally rotating hairy black holes in three-dimensions by using the canonical energy method.

\section{Review: quasi-local formalism for charges}
In this section, we review the quasi-local ADT  formalism based on the identically conserved current and summarize some  properties of the generalized ADT current~\cite{Kim:2013zha,Kim:2013cor,Hyun:2014kfa,Hyun:2014sha,Hyun:2014nma,Peng:2014gha}, which relies on EOM or Euler-Lagrange expression.
To summarize our conventions and present results succinctly, let us denote the metric and matter fields collectively as $\Psi=(g,\psi)$. The Euler-Lagrange expression can also be written collectively as $\CE_{\Psi} = (\CE_{\mu\nu}, \CE_{\psi})$.
The Bianchi or Noether identity for a diffeomorphism parameter $\zeta^{\mu}$ may be written as
\begin{equation} \label{genBI}
\nabla_{\mu}(2{\bf E}^{\mu\nu}\zeta_{\nu}) = -\CE_{\Psi}\Lie_{\zeta}\Psi\,,
\end{equation}
where $\Lie_{\zeta}$ denotes the Lie derivative along the vector field $\zeta^{\mu}$ and ${\bf E}^{\mu\nu} \equiv \CE^{\mu\nu} - \frac{1}{2}\CZ^{\mu\nu}$. Here,   the $\CZ$-tensor  is given by a linear combination of the product of the matter field $\psi$ and the matter Euler-Lagrange expression $\CE_{\psi}$. Concretely,  for an $\ell$-th rank tensor field $\psi_{\mu_{1} \cdots \mu_{\ell}}$, one can see that the $\CZ$-tensor is given by
\begin{equation} \label{}
 \CZ^{\mu\nu}(\CE_{\psi},\psi) = \CE^{\mu\alpha_{2}\alpha_{3}\cdots\alpha_{\ell}}_{\psi}\, \psi^{\nu}_{~\alpha_{2}\alpha_{3}\cdots\alpha_{\ell}} + \CE^{\alpha_{1}\mu\alpha_{3}\cdots\alpha_{\ell}}_{\psi}\, \psi^{~~\nu}_{\alpha_{1}~\alpha_{3} \cdots\alpha_{\ell}} + \cdots \,,
\end{equation}
which will be represented schematically as $\CZ^{\mu\nu} = \CE^{\mu}_{\psi} \circ \psi^{\nu}$. Note that the $\CZ$-tensor vanishes for minimally coupled scalar fields. 

In the quasi-local ADT formalism, the on-shell ADT current~\cite{Abbott:1981ff,Abbott:1982jh,Deser:2002rt,Deser:2002jk} is generalized to the off-shell level. The off-shell conserved current ${\bf J}^{\mu}$  is composed of two pieces, the generalized off-shell ADT current $\CJ^{\mu}_{ADT}$  and the additional term $\CJ^{\mu}_{\Delta}$, as
\begin{equation} \label{DefOffCur}
{\bf J}^{\mu}  (\zeta\,;\, \Psi, \delta \Psi)= \CJ^{\mu}_{ADT}  (\zeta\,;\, \Psi, \delta \Psi)+ \CJ^{\mu}_{\Delta} (\Psi\,|\,\Lie_{\zeta} \Psi, \delta \Psi)\,,
\end{equation}
where the ADT current is defined by (see~\cite{Hyun:2014kfa,Hyun:2014sha} for some details) 
\begin{equation} \label{DefADT}
\sqrt{-g}\CJ^{\mu}_{ADT} (\zeta\,;\, \Psi, \delta \Psi) = \delta (\sqrt{-g}{\bf E}^{\mu\nu}\zeta_{\nu}) - \sqrt{-g}{\bf E}^{\mu}_{~\nu}\delta\zeta^{\nu} + \frac{1}{2}\sqrt{-g}\zeta^{\mu}\CE_{\Psi}\delta \Psi \,.
\end{equation}
 The additional term ${\cal J}^{\mu}_{\Delta}$, which vanishes when $\zeta$ is a Killing vector,  is introduced to  preserve the off-shell conservation property of the current even for an asymptotic Killing vector $\zeta^{\mu}$. This additional term is given by the iterative integration by parts on a specific combination of the Euler-Lagrange expression~\cite{Hyun:2014kfa} and the current ${\bf J}^{\mu}$ can be shown to be conserved at the off-shell level (see Appendix A for its derivation).  The additional current  ${\cal J}^{\mu}_{\Delta}$ is symplectic and  vanishes for a Killing vector $\zeta^{\mu}$, which can  be related to  the symplectic current $\omega^{\mu} $ in the covariant phase space~\cite{Lee:1990nz} as given in Eq.~(\ref{AddCur}). Despite its relation to the symplectic current, we would like to emphasize that the additional current ${\cal J}^{\mu}_{\Delta}$ is constructed solely from the Euler-Lagrange expression.

Since the off-shell current ${\bf J}^{\mu}$ is conserved identically, it can be written in terms of the (anti-symmetric) off-shell potential ${\bf Q}^{\mu\nu}$ as
\begin{equation} \label{ADTcur}
\sqrt{-g}{\bf J}^{\mu} =\p_{\nu}(\sqrt{-g}{\bf Q}^{\mu\nu})\,,
\end{equation}
and 
the linearized quasi-local ADT charges for an asymptotic Killing vector $\zeta^{\mu}$ are defined by
\begin{equation} \label{ADTcharge}
\delta Q (\zeta) = \frac{1}{8\pi G}\int dx_{\mu\nu}\sqrt{-g}{\bf Q}^{\mu\nu}(\zeta\,;\, \Psi,\delta\Psi)\,,
\end{equation}
where this expression should be evaluated on-shell at the final stage of computation. Here, we would like to emphasize that the off-shell current ${\bf J}^{\mu}$ and potential ${\bf Q}^{\mu\nu}$ are constructed from the Euler-Lagrange expression and thus they are free from the ambiguity or the non-covariance in the Lagrangian.
As is derived in Appendix A, the off-shell potential ${\bf Q}^{\mu\nu}$ can be related to the Noether potential $K^{\mu\nu}$ and the surface term $\Theta^{\mu}$ of the Lagrangian, up to a total derivative term, as
\begin{equation} \label{}
2\sqrt{-g}{\bf Q}^{\mu\nu}(\zeta\,;\,\Psi,\delta \Psi) = \delta K^{\mu\nu}(\zeta)  - K^{\mu\nu}(\delta \zeta) -2\zeta^{[\mu}\Theta^{\nu]}(\delta \Psi)+ \sqrt{-g}{\bf A}^{\mu\nu}(\Psi\,|\,\Lie_{\zeta}\Psi,\delta \Psi)\,.
\end{equation}
For a Killing vector, the charge expression is completely consistent with the conventional ADT expression at the asymptotic infinity and is also consistent with the covariant phase space formalism, for instance, in the computation of the black hole entropy. Furthermore, for asymptotic Killing vectors, this charge expression can be used to obtain asymptotic symmetry generators\cite{Hyun:2014kfa}.

Before going ahead, we would like to remark the special properties of the ADT current $\CJ^{\mu}_{ADT}$ and the additional current $\CJ^{\mu}_{\Delta}$. Firstly, 
one may note that the ADT current, $\CJ^{\mu}_{ADT}(\zeta\,;\, \Psi,\delta \Psi)$, is off-shell conserved when $\zeta$ is a Killing vector, and that it  depends linearly  on $\zeta$.  This ADT current can be written in terms of a differential operator ${\cal D}_{\Psi}$ acting on the field variation $\delta\Psi$ as
\begin{equation} \label{DifADTcur}
\CJ^{\mu}_{ADT} (\zeta\,;\, \Psi, \delta \Psi)  = \zeta_{\nu}({\cal D}_{\Psi}\delta \Psi)^{\mu\nu} \,.
\end{equation}
When the background configuration $\Psi$ satisfies EOM, $\CE_{\Psi}=0$, the current  reduces to
\begin{equation} \label{}
\CJ^{\mu}_{ADT} (\zeta\,;\, \Psi, \delta \Psi) \Big|_{\CE_{\Psi}=0} = \delta {\bf E}^{\mu\nu}\, \zeta_{\nu}=  \Big(\delta\CE^{\mu\nu} - \frac{1}{2}\delta\CE^{\mu}_{\psi}\circ \psi^{\nu}\Big)\zeta_{\nu}\,,
\end{equation}
which vanishes for an arbitrary $\zeta^{\mu}$ when we impose that  $\delta \Psi$ also satisfies  linearized EOM, $\delta \CE_{\Psi}=0$. This property will be called the ``on-shell'' vanishing property of the ADT current ${\cal J}^{\mu}_{ADT}$ in the following.  In summary, the  generalized off-shell ADT current ${\cal J}^{\mu}_{ADT}$  has the off-shell conservation property for a Killing vector and the on-shell vanishing property for an arbitrary $\zeta$ as
\begin{align}   \label{}
\nabla_{\mu}{\cal J}^{\mu}_{ADT}(\zeta\,;\, \Psi, \delta \Psi)  &=0  \qquad~~~   {\rm off~shell} \qquad {\rm for }~~ {\rm  a~Killing~ vector}~~ \zeta\,,   \\
{\cal J}^{\mu}_{ADT}(\zeta\,;\, \Psi, \delta \Psi)  &= 0 \qquad \quad   {\rm on~ shell}  \qquad   {\rm for}~~ {\rm arbitrary}~~ \zeta \,.  \label{osvan}
\end{align}
Since ${\cal J}^{\mu}_{\Delta}=0$ for a Killing vector $K$, 
the above off-shell conservation property for a Killing vector  can be rephrased as 
\begin{equation} \label{}
{\cal J}^{\mu}_{ADT}(K) = {\bf J}^{\mu}(K) = \nabla_{\nu}{\bf Q}^{\mu\nu}(K)\,. \nn
\end{equation}

Now,  let us consider  the  second variation of the ADT current for an arbitrary vector $\zeta^{\mu}$, which would be relevant for the construction of the bilinear form on the first order variation space. Note that the variation of the ADT current $\CJ^{\mu}_{ADT}(\zeta\,;\, \Psi, \delta_{1}\Psi)$ can be written in terms of three pieces as
\begin{equation} \label{}
\delta_{2}\Big( \CJ^{\mu}_{ADT}(\zeta\,;\,\Psi, \delta_{1}\Psi)\Big) =\CJ^{\mu}_{ADT}(\delta_{2}\zeta\,;\, \Psi, \delta_{1}\Psi)  +\CJ^{\mu}_{ADT}(\zeta\,;\,\delta_{2}\Psi, \delta_{1}\Psi) + \CJ^{\mu}_{ADT}(\zeta\,;\,\Psi, \delta_{2}\delta_{1}\Psi)  \,,
\end{equation}
%
where the first and the last term in the right-hand side are given through the representation of the ADT current  $\CJ^{\mu}_{ADT}$  in Eq.~(\ref{DifADTcur}), explicitly by
\begin{equation} \label{}
\CJ^{\mu}_{ADT}(\delta_{2}\zeta\,;\, \Psi, \delta_{1}\Psi) = \delta_{2}\zeta^{\nu}({\cal D}_{\Psi}\, \delta_{1}\Psi)^{\mu}_{~\nu}\,, \qquad \CJ^{\mu}_{ADT}(\zeta\,;\,\Psi, \delta_{2}\delta_{1}\Psi) = \zeta^{\nu}({\cal D}_{\Psi}\, \delta_{2}\delta_{1}\Psi)^{\mu}_{~\nu}\,.
\end{equation}
Therefore the second term in the right-hand side may be thought of as defined by the variations of the ADT current  as 
\begin{equation} \label{SecVar}
\CJ^{\mu}_{ADT}(\zeta\,;\,\delta_{2}\Psi, \delta_{1}\Psi) \equiv \delta_{2}\Big( \CJ^{\mu}_{ADT}(\zeta\,;\,\Psi, \delta_{1}\Psi)\Big)  - \CJ^{\mu}_{ADT}(\delta_{2}\zeta\,;\, \Psi, \delta_{1}\Psi) -\CJ^{\mu}_{ADT}(\zeta\,;\,\Psi, \delta_{2}\delta_{1}\Psi) \,,
\end{equation}
which would be a good candidate for a symmetric bilinear form on the space of on-shell first order variations or linear perturbations. 
Note also that this current expression $\CJ^{\mu}_{ADT}(\zeta\,;\, \delta_{2}\Psi,\delta_{1}\Psi)$  depends linearly on the vector $\zeta^{\mu}$ while it is independent of the derivatives of $\zeta^{\mu}$, as can be inferred from the representation of  the ADT current given in Eq.~(\ref{DifADTcur}).

\section{Modified canonical energy}
%

By using the ADT current $\CJ^{\mu}_{ADT}$, we would like to introduce a modified canonical energy $\CE(\delta_{1}\Psi, \delta_{2}\Psi)$,  whose original form was defined in~\cite{Hollands:2012sf}.  The essential properties of the canonical energy proposed in~\cite{Hollands:2012sf} may be summarized as follows. It is   a symmetric bilinear 
form on the first field variation $\delta \Psi$, gauge-invariant,  monotonic along the ``time evolution'' and conserved in the sense that it does not depend on the choice of the  Cauchy surface for given boundaries. As will be shown in the following, one may use the ADT current $\CJ^{\mu}_{ADT}$ instead of the symplectic current $\omega^{\mu}$ to construct a canonical energy with the alluded properties.  

Let us  denote the Killing vector for the background $\Psi$ as $K$. The modified canonical energy for an exact  Killing vector $K$ can be introduced through the second variation of the ADT current by\footnote{In the context of the stability of black holes or branes~\cite{Hollands:2012sf,Hollands:2014lra},  we need to impose the axisymmetric condition for linear perturbations and the Killing vector $K$  can be chosen as the stationary one $K_{T}=\frac{\p}{\p t}$ in the asymptotically flat case. In the asymptotically AdS case, we can take the Killing vector $K$ as the horizon Killing vector $K_{H}$ without imposing the axisymmetric conditions.}
\begin{equation} \label{CanE}
\CE(K\,;\, \delta_{1} \Psi,\delta_{2}\Psi) \equiv  -\frac{1}{8\pi G}\int_{\Sigma}dx_{\mu}~ 
 \sqrt{-g} \CJ^{\mu}_{ADT}(K\,;\,\delta_{2}\Psi, \delta_{1}\Psi) \,,
\end{equation}
which is  different, at least  apparently,  from the definition given in Ref.~\cite{Hollands:2012sf}. Here, $\Sigma$ denotes a Cauchy surface extending from the bifurcation surface $B$ to the spacelike infinity. 
Nevertheless, the essential features for the canonical energy will be shown to be satisfied. To this purpose, let us look into the properties of the  current expression $\CJ^{\mu}_{ADT}(K\,;\, \delta_{2}\Psi,\delta_{1}\Psi)$: symmetric form, conservation, gauge invariance and the monotonicity. In the following, we take K as the horizon Killing vector whenever the choice is convenient to present. \\

{\bf Symmetric bilinear form}: 
By using EOM, $\CE_{\Psi}=0$ and linearized EOM, $\delta \CE_{\Psi}=0$ with the interchangeability of two generic variations $\delta_{1}$ and $\delta_{2}$, such as $\delta_{2}\delta_{1}\Psi=\delta_{1}\delta_{2}\Psi$ and  $\delta_{2}\delta_{1}\CE_{\Psi}= \delta_{1}\delta_{2}\CE_{\Psi}$, one can see that the second variations of the ADT current satisfy the relation as\footnote{Here, `on-shell' means that  EOM $\CE_{\Psi}=0$ and linearized EOM $\delta \CE_{\Psi}=0$ are satisfied without requiring the second order variation of   EOM  to be satisfied.}.
\begin{equation} \label{}
\delta_{2}\Big( \CJ^{\mu}_{ADT}(\zeta\,;\,\Psi, \delta_{1}\Psi)\Big)\Big|_{on-shell} = (\delta_{2}\delta_{1}{\bf E}^{\mu}_{~\nu})\zeta^{\nu}|_{on-shell} = \delta_{1}\Big( \CJ^{\mu}_{ADT}(\zeta\,;\,\Psi, \delta_{2}\Psi)\Big)\Big|_{on-shell}\,,
\end{equation}
where we have used the definition of the (off-shell) ADT current given in Eq.~(\ref{DefADT}).
Because of the interchangeability of two generic variations,  $\CJ^{\mu}_{ADT}(\zeta\,;\, \Psi,\delta_{2}\delta_{1}\Psi)$ is also symmetric over two variations, $\delta_{1}$ and $\delta_{2}$. 
Thus, by using the defining relation for $\CJ^{\mu}_{ADT}(\zeta\,;\, \delta_{2}\Psi,\delta_{1}\Psi)$ in Eq.~(\ref{SecVar}) and by using the `on-shell' vanishing property of the ADT current $\CJ^{\mu}_{ADT}(\delta_{2}\zeta\,;\, \Psi, \delta_{1}\Psi)$ in Eq.~(\ref{osvan}), one can see that the current expression $\CJ^{\mu}_{ADT}(\zeta\,;\, \delta_{2}\Psi, \delta_{1}\Psi)$ is a symmetric bilinear form on the space of  on-shell first order variations for an arbitrary vector $\zeta^{\mu}$, while  it is independent of the second order field variations.  \\

{\bf Conservation}: 
Recall that, for the exact Killing vector   $K$ for the arbitrary background  $\Psi$ with the arbitrary $\delta \Psi$,  the additional current $\CJ^{\mu}_{\Delta}(\Lie_{K}\Psi,\delta\Psi)$ vanishes and so the off-shell current ${\bf J}^{\mu}(K\,;\, \Psi, \delta \Psi)$ reduces to  the ADT current $\CJ^{\mu}_{ADT}(K\,;\, \Psi, \delta \Psi)$. Thus, just by replacing the arbitrary $\delta \Psi$ with $\delta_{2}\delta_{1}\Psi$ in Eq.~(\ref{ADTcur}), one obtains
\begin{equation} \label{Rel1}
\sqrt{-g} \CJ^{\mu}_{ADT}(K\,;\, \Psi, \delta_{2}\delta_{1} \Psi)  =\p_{\nu}\Big(\sqrt{-g}\,  {\bf Q}^{\mu\nu}(K\,;\, \Psi, \delta_{2}\delta_{1}\Psi ) \Big)\,,
\end{equation}
%
 which shows us  that the last term in the right-hand side of Eq.~(\ref{SecVar}) is identically conserved in the Killing vector case. The second term in the right-hand side of Eq.~(\ref{SecVar})  vanishes identically when the `on-shell' condition is imposed.
By using the Killing property of the background $\Lie_{K}\Psi=0$ and using Eq.~(\ref{conA}), one can see  that the variation of the ADT current for the Killing vector $K$ is also  ``on-shell'' conserved: 
\begin{equation} \label{Rel2}
\p_{\mu}\Big[\delta_{2}\Big( \sqrt{-g}\CJ^{\mu}_{ADT}(K\,;\,\Psi, \delta_{1}\Psi)\Big)\Big]_{on-shell}=  \sqrt{-g}~(\delta_{2}\delta_{1}\CE_{\Psi})\Lie_{K}\Psi\Big|_{on-shell}=0\,,
\end{equation}
where the on-shell vanishing property of the current $\CJ^{\mu}_{ADT}(K\,;\,\Psi,\delta\Psi)$ is also used. Note that   the  variation of the ADT current $\delta (\sqrt{-g}\CJ^{\mu}_{ADT}(K\,;\,\Psi, \delta\Psi))$  is conserved without the requirement that  the second order variation of field $\Psi$ satisfy EOM\footnote{When the second order variation of the field $\Psi$ also satisfies EOM,   the variation of the ADT current itself vanishes, {\it i.e.} $\delta(\sqrt{-g}\CJ^{\mu}_{ADT}(\delta \Psi)) =0$.}. Collecting the above conservation and/or vanishing properties of  each term in the right-hand side in Eq.~(\ref{SecVar}), one can see that the current expression $\CJ^{\mu}_{ADT}(K\,;\, \delta_{2}\Psi, \delta_{1}\Psi)$ is conserved at the ``on-shell'' level.\\    

 {\bf Gauge invariance}: The previously established two properties of the current expression, $\CJ^{\mu}_{ADT}(K\,;\, \delta_{2}\Psi,\delta_{1}\Psi)$,  motivate the introduction of the modified canonical energy $\CE(K\,;\,\delta_{1}\Psi,\delta_{2}\Psi)$ for a Killing vector $K$ in Eq.~(\ref{CanE}).   To see the diffeomorphism transformation property of the current expression $\CJ^{\mu}_{ADT}(K\,;\, \delta_{2}\Psi,\delta_{1}\Psi)$ and the gauge invariance of the modified canonical energy,  note that the Lie derivative of the ADT current  can be written as
\begin{equation} \label{Gauge1}
\Lie_{\epsilon}\CJ^{\mu}_{ADT}(K\,;\, \Psi,\delta \Psi)  = \CJ^{\mu}_{ADT}(\Lie_{\epsilon} K \,;\, \Psi, \delta \Psi) +  \CJ^{\mu}_{ADT}(K\,;\,\Lie_{\epsilon}\Psi, \delta \Psi) + \CJ^{\mu}_{ADT}(K\,;\,\Psi, \Lie_{\epsilon}\delta\Psi)\,,
\end{equation}
where the first term in the right-hand side vanishes  when the  ``on-shell''  conditions are imposed  because of the ``on-shell'' vanishing property  of $\CJ^{\mu}_{ADT}(\zeta\,;\, \Psi,\delta\Psi)$ for an arbitrary $\zeta^{\mu}$.  One may  note that the Lie derivative of the ADT current can also be written as
\begin{align}   \label{Gauge2}
\Lie_{\epsilon}\CJ^{\mu}_{ADT}(K\,;\, \Psi,\delta \Psi)  &= \nabla_{\nu}\Big(2\epsilon^{[\nu}\CJ^{\mu]}_{ADT}(K\,;\, \Psi,\delta \Psi) \Big) \nn \\
&\quad -\CJ^{\mu}_{ADT}(K\,;\, \Psi,\delta \Psi)\nabla_{\nu}\epsilon^{\nu}+\epsilon^{\mu}\nabla_{\nu}\CJ^{\nu}_{ADT}(K\,;\, \Psi,\delta \Psi)\,, 
\end{align}
where the second term in the right-hand side vanishes when   the ``on-shell'' conditions are  imposed and the last term vanishes because of the conservation property of the ADT current $\CJ^{\mu}_{ADT}(K\,;\, \Psi,\delta \Psi)$  for the Killing vector $K$.  Combining two expressions in Eq.~(\ref{Gauge1}) and~(\ref{Gauge2}), one can show that
\begin{align}   \label{}
&- \int dx_{\mu}\sqrt{-g}\CJ^{\mu}_{ADT}(K\,;\,\Lie_{\epsilon}\Psi, \delta \Psi)|_{on-shell}  \nn \\
&\qquad = \int dx_{\mu}\sqrt{-g} \Big[\CJ^{\mu}_{ADT}(K\,;\,\Psi, \Lie_{\epsilon}\delta\Psi) -\Lie_{\epsilon}\CJ^{\mu}_{ADT}(K\,;\, \Psi,\delta \Psi)\Big]_{on-shell} \nn\\
&\qquad  = \int dx_{\mu\nu}\sqrt{-g}\Big[{\bf Q}^{\mu\nu}(K\,;\,\Psi,\Lie_{\epsilon}\delta\Psi)  + 2\epsilon^{[\mu}\CJ^{\nu]}_{ADT}(K\,;\,\Psi,\delta\Psi) \Big]_{on-shell} \,,  \nn
\end{align}
where we have used $\CJ^{\mu}_{ADT}(K\,;\,\Psi, \Lie_{\epsilon}\delta\Psi) = \nabla_{\nu}{\bf Q}^{\mu\nu}(K\,;\,\Psi,\Lie_{\epsilon}\delta\Psi)$ for the Killing vector $K$. Note also that the last term in the second equality also vanishes by the `on-shell'  vanishing condition of the ADT current ${\cal J}^{\mu}_{ADT}$. In the end, one obtains
\begin{equation} \label{CanEgauge}
\CE(K\,;\, \Lie_{\epsilon} \Psi,\delta\Psi)  = \frac{1}{8\pi G}\int_{\partial \Sigma} dx_{\mu\nu}\sqrt{-g}\, {\bf Q}^{\mu\nu}(K\,;\,\Psi,\Lie_{\epsilon}\delta\Psi)  |_{on-shell}\,,
\end{equation}
which shows us that a generic diffeomorphism transformation leads to a total boundary term and the modified canonical energy becomes  invariant for the local, compactly supported  diffeomorphism parameter $\epsilon^{\mu}$ on $\Sigma$.  

More generically, $\epsilon^{\mu}$ may not be local, compactly supported, which could be present in our setup.  In pure Einstein gravity, as was done in Ref.~\cite{Hollands:2012sf}, one  may choose gauges of the metric near the horizon such that the linear perturbation does not change the expansion of the bifurcation surface $B$ as $\delta \vartheta|_{B}=0$ and $\delta A=0$ for the area $A$ of the surface $B$. Concretely, on the near horizon, one can take the metric in the Gaussian null coordinates~\cite{Moncrief:1983,Hollands:2006rj,Hollands:2012sf}  generically as
\begin{equation} \label{GaussN}
ds^{2}_{NH} = 2du(dr - r^{2}\alpha du - r\beta_{a}dx^{a}) +\mu_{ab}dx^{a}dx^{b}\,,
\end{equation}
where $u$ and $r$ correspond to affine parameters  along the null directions and $\mu_{\alpha\beta}$ denotes the metric on the sphere part in Gaussian null coordinates. Here,  two null coordinate vectors  $n \equiv n^{\mu}\partial_{\mu} =\frac{\partial}{\partial u}$ and $\ell\equiv \ell^{\mu}\partial_{\mu} = \frac{\partial}{\partial r}$ commute  and $n^{\mu}$ is normal to the horizon with the relation $n^{\mu}\ell_{\mu}=1$. In the chosen gauge,  the horizon Killing vector field for the background near the horizon is given by
\begin{equation} \label{}
 K_{H}= \kappa\Big(u\frac{\partial }{\partial u} - r \frac{\partial }{\partial r}\Big)\,,
\end{equation}
where $\kappa$ denotes the surface gravity 
and the perturbed metric becomes 
\begin{equation} \label{PertH}
h_{\mu\nu}dx^{\mu}dx^{\nu} = -2r^{2}\delta \alpha\, du^{2} -2r\delta \beta_{a}\,dx^{a}du + \delta \mu_{ab}dx^{a}dx^{b}\,,
\end{equation}
where $h_{\mu\nu} \equiv \delta g_{\mu\nu}$. At  infinity in the asymptotic flat case, the ``Bondi gauge'' is chosen with a further choice as was done by Geroch-Xanthopoulos~\cite{Geroch:1978ur,Hollands:2006rj}, where the ``unphysical metric'' at  infinity may be taken as
\begin{equation} \label{BondiGauge}
d\tilde{s}^{2} = \tilde{\Omega}^{2}ds^{2} = 2d\tilde{\Omega}d\tilde{u} + \tilde{\mu}_{ab}d\tilde{x}^{a}d\tilde{x}^{b} + {\cal O}(\tilde{\Omega})\,,
\end{equation}
and the linearized metric satisfies $\tilde{g}^{\mu\nu}\tilde{h}_{\mu\nu} = {\cal O}(\tilde{\Omega})$. In the case of the asymptotically AdS boundary conditions, the unphysical metric is taken as~\cite{Hollands:2005wt,Green:2015kur}
\begin{equation} \label{}
d\tilde{s}^{2} = d\tilde{\Omega}^{2} + \tilde{\gamma}_{\mu\nu}dx^{\mu}dx^{\nu}+{\cal O}(\tilde{\Omega}^{2})\,,
\end{equation}
where $\tilde{\gamma}_{\mu\nu}$ denotes the metric on the boundary, {\it i.e.} the metric of the Einstein static universe. 

And then,  further conditions are imposed such that  the  charges $Q(\zeta)$ for the asymptotic symmetry generator $\zeta$ are not changed under linear perturbations. This imposition is related to the perturbation {\it toward stationary black holes}, which would generate unwanted contributions~\cite{Hollands:2012sf}.  There are still remnant gauge transformations preserving the chosen gauge conditions, of which  gauge parameter  $\epsilon^{\mu}$ becomes tangent to the horizon  and corresponds to the asymptotic symmetry generators near  infinity.   For such a noncompact gauge parameter $\epsilon^{\mu}$, one can show the gauge invariance of the modified canonical energy by using the results given in Appendix B. To phrase it simply,  under the same  assumptions as in Ref.~\cite{Hollands:2012sf} on the horizon and asymptotic behavior of the gauge parameter $\epsilon$, one can argue that  
\begin{align}   \label{CanEgauge2}
\CE(K\,;\, \Lie_{\epsilon} \Psi,\delta\Psi)  = \frac{1}{8\pi G}\Big(\int_{\infty} -\int_{B}\Big)dx_{\mu\nu}\sqrt{-g}\, {\bf Q}^{\mu\nu}(K\,;\,\Psi,\Lie_{ \epsilon}\delta\Psi)  |_{on-shell}= 0\,.
\end{align}
In summary, the modified canonical energy is also gauge invariant  when the appropriate conditions are taken.\\

%
%

{\bf Monotonic property}:  To consider the monotonic property along the ``time evolution''  of the canonical energy, Hollands {\it et al.} evaluated the  canonical energy on the four sectors\footnote{See the second figure in Ref.~\cite{Hollands:2012sf} and figure 2 in Ref.~\cite{Hollands:2014lra}.} ${\mathscr I}(t_{1}) \cup {\mathscr I}(t_{2}) \cup {\mathscr H}_{12}\cup {\mathscr J}_{12}$ with $t_{1}< t_{2}$, where ${\mathscr H}_{12}$ and ${\mathscr J}_{12}$ denote the regions in the future horizon and future null/spacelike infinity, respectively. In order to obtain rigorous statements about the behavior of the canonical energy, some machinery is utilized for taking care of the null infinity and the horizon. Briefly speaking, one needs  to choose appropriate gauges and falloff conditions~\cite{Hollands:2012sf,Hollands:2014lra,Green:2015kur}. In the asymptotic flat spacetime, the null infinity should be managed carefully and the statement is proven only for the even-dimensional case. As was alluded before,  $K$ is taken  as the stationary  Killing vector $K_{T}$ in this case.   To the contrary, in the asymptotic AdS spacetime, the asymptotic infinity is timelike and flux cannot leak away.  In this case, the contribution from the asymptotic infinity is trivial and there is no restriction on the dimensionality.

Instead of performing this analysis directly in our construction, we would like to relate our  modified canonical energy to the original expression of the canonical energy given in~\cite{Hollands:2012sf} and borrow the monotonicity property of the HW canonical energy to show the monotonicity of the modified canonical energy in our construction. 
Recall that the symplectic form $W_{\Sigma}$ in the covariant phase space is defined by 
\begin{equation} \label{SympF}
W_{\Sigma}(\Psi\,|\, \delta_{1}\Psi,\delta_{2} \Psi) \equiv \frac{1}{16\pi G}\int_{\Sigma} dx_{\mu}~ \omega^{\mu} (\Psi\,|\, \delta_{1}\Psi,\delta_{2}\Psi)\Big|_{on-shell} \,,
\end{equation}
and that  the HW  canonical energy for the Killing vector $K$ is defined by
\begin{equation} \label{}
\CE_{HW}(K\,;\, \delta_{1} \Psi,\delta_{2}\Psi) \equiv
 W_{\Sigma}(\Psi\,|\, \Lie_{K}\delta_{2}\Psi,\delta_{1} \Psi)\,,
\end{equation}
where $\delta K^{\mu}=0$ is assumed as before for simplicity. Therefore, the difference between our modified canonical energy and its expression in the HW canonical energy turns out  to be just  surface terms, as is shown in Appendix C. Explicitly, the difference is given by
\begin{align}   \label{RelHW}
&\CE_{HW}(K\,;\, \delta\Psi,\delta\Psi)- \CE (K\,;\, \delta\Psi,\delta\Psi)  
\nn \\
&\quad = \frac{1}{16\pi G} \int_{\partial \Sigma}dx_{\mu\nu}~ \bigg[2\sqrt{-g}{\bf Q}^{\mu\nu}(K\,;\,\delta\Psi, \delta\Psi)   - \sqrt{-g}{\bf A}^{\mu\nu}(\Psi\,|\,\Lie_{K}\delta\Psi, \delta\Psi) \bigg]_{on-shell}~,
\end{align}
%
%
%
where the boundary $\partial \Sigma$ is composed of two parts, $\int_{\partial \Sigma} = \int_{\infty} - \int_{B}$. Basically, these boundaries are sphere parts of geometry on the horizon and the infinity. The contribution from the infinity vanishes as can be inferred from the linearized charge expression in Eq.~(\ref{ADTcharge}), which is finite and  taken to vanish for  linear perturbations. 
In pure Einstein gravity, one can infer from Eq.~(\ref{ReltoHW1}) that the above boundary term at $B(t)$ is given by\footnote{The similar expression at $B(t=0)$ was obtained in  Eq.~(85) in Ref.~\cite{Hollands:2012sf}.}
\begin{align}   \label{Bterm}
\int_{B(t)} &\equiv   \frac{1}{8\pi G} \int_{B(t)}dx_{\mu\nu}\, \sqrt{-g}{\bf Q}^{\mu\nu}(K_H\,;\,\delta g, \delta g)  \Big|_{on-shell}   \nn \\
&=\frac{1}{8\pi G} \int_{B(t)} dx \sqrt{\mu}~ \Big[\kappa\,\delta\mu^{\alpha\beta}\delta \mu_{\alpha\beta} - \frac{3}{2} \delta \mu^{\alpha\beta}\Lie_{K_{H}}\delta \mu_{\alpha\beta}\Big]_{on-shell}  \,. 
\end{align}
Note that the second term in the right-hand side of the last equality vanish on the bifurcation surface $B(t=0)$ because of $K_{H}\rightarrow 0$, while it would vanishes at $B(t)$ when the perturbed shear vanishes as was argued in Ref.~\cite{Hollands:2012sf}.

The absence of the contribution from the infinity implies that there would no difference between ${\cal E}$ and ${\cal E}_{HW}$ on the sector ${\mathscr J}_{12}$, while they may be different on other sectors ${\mathscr I}(t_{1}), {\mathscr I}(t_{2})$, and ${\mathscr H}_{12}$ up to boundary terms given by the integral over $B(t_{1})$ and $B(t_{2})$. Schematically, the difference on these sectors can be written as
\begin{equation} \label{}
\CE_{HW}\Big|_{{\mathscr I}(t_{1,2})} = \CE \Big|_{{\mathscr I}(t_{1,2})} -\int_{B(t_{1,2})}\,, \qquad \CE_{HW}\Big|_{{\mathscr H}_{12}} = \CE \Big|_{{\mathscr H}_{12}} + \int_{B(t_{2})}- \int_{B(t_{1})}\,,
\end{equation}
where we have abused the notation to denote the symplectic form and its counterpart on the sector ${\mathscr H}_{12}$ as ${\cal E}_{HW}$ and ${\cal E}$, respectively.
This is consistent with the individual conservation of ${\cal E}_{HW}$ and ${\cal E}$.
%
%


To show the monotonicity of the canonical energy in Ref.~\cite{Hollands:2012sf}, the appropriate boundary terms are subtracted from the canonical energy  to modify ${\cal E}_{HW}$ as  $\bar{\cal E}_{HW}$ and then it was shown that this barred canonical energy satisfies for $t_{1} \le t_{2}$
\begin{equation} \label{Mono}
\bar{\cal E}_{HW} (t_{2}) - \bar{\cal E}_{HW} (t_{1})  \le 0\,.
\end{equation}
Simply by defining  our barred quantity as $\bar{\cal  E} (t)= \bar{\cal E}_{HW}(t)$, one can establish  immediately the monotonicity property of our barred canonical energy. The actual application of the canonical energy comes from the fact that $\bar{\cal E}_{HW}(t) \rightarrow {\cal E}_{HW}(t=0)$ as $t\rightarrow 0$, which can be used to argue that  ${\cal E}_{HW}(t=0) < 0 $ implies the instability.

In order to see how to use our canonical energy for the stability, 
it is useful to  introduce the boundary term composed solely of the first term in the last equality in Eq.~(\ref{Bterm}) by denoting it as  $\int'_{B(t)}$, which is negative semi-definite. Note that our barred canonical energy is different from the unbarred one even at $t=0$:
\begin{equation} \label{}
\bar{\cal E} = \bar{\cal E}_{HW}  = {\cal E}_{HW} = {\cal E} -\int'_{B} \,,
\end{equation}
where we have used  the relation between ${\cal E}_{HW}$ and ${\cal E}$ and the fact that $\bar{\cal E}_{HW} = {\cal E}_{HW}$ at $t=0$. Because of this relation, $\CE (t=0) <0$ may not imply the instability for a generic non-compact perturbation. 
%
However, we may prepare the initial data at $t=t_{1}$, which are compactly supported such that ${\cal E}(t_{1})=\bar{\cal  E}(t_{1}) = \bar{\cal E}_{HW}(t_{1})$. Then, at later time we can conclude that
\begin{equation} \label{}
\bar{\cal E}(t_{2}) = \bar{\cal E}_{HW}(t_{2})  \le \bar{\cal E}_{HW} (t_{1}) =\bar{\cal E} (t_{1})={\cal E}(t_{1})\,,
\end{equation}
Thus, for the compactly supported initial data,
the instability argument works in our modified canonical energy:  ${\cal E} (t=0) < 0$ implies the instability  just as ${\cal E}_{HW}(t=0)  <  0$ does. \\ 

%

Now, we would like to give various comments on our construction and its meaning.   In our construction the symmetric bilinear property of the modified canonical energy on the first order variations  is manifest even on $\Sigma(t)$, in contrast to the expression of the canonical energy in the HW construction. In the HW construction, the appropriate gauges at the asymptotic infinity and the horizon are used to show this property on $\Sigma(t=0)$, only. 
As can be inferred from Eq.~(\ref{SecVar}) and the property of the ADT current, our modified canonical energy can also be written, when the second order variation of field $\Psi$ satisfies second order EOM,  as 
\begin{equation} \label{}
\CE (K\,;\, \delta \Psi,\delta \Psi)   =   \frac{1}{8\pi G} \int_{\p\Sigma}dx_{\mu\nu}~ \sqrt{-g}{\bf Q}^{\mu\nu}(K\,;\,\Psi, \delta^{2}\Psi)|_{\CE_{\Psi}=\delta\CE_{\Psi}=\delta^{2}\CE_{\Psi}=0}\,,
\end{equation}
which means  that the bulk expression of the modified canonical energy expression becomes the total surface term. More concretely, the second order variation of conserved charges in our construction can be read from Eq.~(\ref{ADTcharge}) as 
\begin{align}   \label{}
\delta^{2} Q(K) &= \frac{1}{8\pi G}\int_{\partial \Sigma} dx_{\mu\nu}~ \delta\Big[\sqrt{-g}{\bf Q}^{\mu\nu}(K\,;\, \Psi,\delta\Psi) \Big] \nn \\
&= \frac{1}{8\pi G}\int_{\partial \Sigma} dx_{\mu\nu}~ \sqrt{-g}\Big[{\bf Q}^{\mu\nu}(K\,;\, \delta\Psi,\delta\Psi) +{\bf Q}^{\mu\nu}(K\,;\, \Psi,\delta^{2}\Psi) \Big]\,,
\end{align}
where we used  $\delta K=0$ and  $\delta\sqrt{-g}=0$ at $\partial\Sigma$. By using the second order perturbation satisfying  EOM and falloff conditions of $\delta \Psi$, one can see that
\begin{align}   \label{2ndCan}
\CE(K_{H}\,;\, \delta \Psi,\delta\Psi) 
&=\delta^{2}M_{\infty}- \Omega_{H}\delta^{2} J_{\infty} - \frac{\kappa}{2\pi}\delta^{2} \CS_{BH}  - \frac{1}{8\pi G}\int_{B} dx_{\mu\nu}~ \sqrt{-g}\, {\bf Q}^{\mu\nu}(K_{H}\,;\, \delta\Psi,\delta\Psi)\,, 
\end{align}
where the last term in pure Einstein gravity is given by Eq.~(\ref{ReltoHW1}).  

On the sector ${\mathscr J}_{12}$ in Einstein gravity without matter fields, one may obtain directly the expression of our modified canonical energy, which is nothing but the second order Einstein tensor  in this case, by choosing the Geroch-Xanthopoulos gauge with additional falloff conditions as in~\cite{Habisohn:1986}. The final expression given in Eq.(4.13) in Ref.~\cite{Habisohn:1986} shows us that our modified energy also gives the same expression as the original canonical energy on this sector.  Therefore, there is no difference between ${\cal E}_{HW}$ and ${\cal E}$ on the sector ${\mathscr J}_{12}$ in Einstein gravity without matter fields, indeed. This direct computation reinforces our argument for the absence of the difference at the infinity between  our modified canonical energy and the HW canonical energy.  One may perform the similar direct computation on ${\mathscr H}_{12}$, since the expressions on ${\mathscr H}_{12}$ and  ${\mathscr J}_{12}$ may be parallel. In the above, the difference is indirectly   shown to reside only on the surface $B(t_{1,2})$ in Appendix C.

Though our expression of linearized conserved charges is completely consistent with the one in the covariant phase space approach and the one from the conventional ADT formalism, the second order variation of conserved charges in our construction may be different from the one in the covariant phase space approach since 
\begin{align}
\delta_{2}\delta_{1}Q(K) &= \delta_{2}\delta_{1}Q_{cov}(K) + \frac{1}{16\pi G}\int_{\partial \Sigma} dx_{\mu\nu}~ \sqrt{-g}{\bf A}^{\mu\nu}(\Lie_{K}\delta_{2}\Psi, \delta_{1}\Psi)\,,
\end{align}
where $Q_{cov}$ denotes the charge in the covariant phase space approach and  we have used the relation given in Eq.~(\ref{ADTpot}) with the condition $\delta K^{\mu}=0$. Indeed, in some higher derivative gravity it was noticed that the additional contribution to the covariant phase space charge expression is important in the context of the Kerr/CFT correspondence~\cite{Azeyanagi:2009wf}. The combination in the second order variations in the canonical energy may also be affected by this difference. This is reflected in the following representation of the HW canonical energy  
\begin{align}   \label{}
\CE_{HW}(K\,;\, \delta \Psi,\delta\Psi)  =\delta^{2}M^{cov}_{\infty}- \Omega_{H}\delta^{2} J^{cov}_{\infty} - \frac{\kappa}{2\pi}\delta^{2} \CS^{cov}_{BH}\,, \nn
\end{align}
which is consistent with the difference given in Eq.~(\ref{RelHW}) and Eq.~(\ref{2ndCan}). 
%

 Practically, the modified canonical energy can be obtained simply by keeping the first order variation terms in the expression of $\delta_{2}\delta_{1} {\bf E}^{\mu\nu}$. 
By using EOM $\CE_{\Psi}=0$ and  linearized EOM $\delta \CE_{\Psi}=0$, respectively, one can see that
\begin{equation} \label{}
  \delta_{2}\Big(\sqrt{-g}\CJ^{\mu}_{ADT} (K\,;\, \Psi,\delta_{1}\Psi)\Big) = \sqrt{-g}(\delta_{2}\delta_{1}{\bf E}^{\mu\nu})K_{\nu}\Big|_{on-shell}\,.
\end{equation}
As a result, the modified canonical energy is given by
\begin{equation} \label{modCE}
\CE(K\,;\, \delta_{1} \Psi,\delta_{2}\Psi) \equiv  -\frac{1}{8\pi G}\int_{\Sigma}dx_{\mu}~ \sqrt{-g}(\delta_{2}\delta_{1}{\bf E}^{\mu\nu})K_{\nu}\Big|^{\delta^{2}\Psi=0}_{on-shell} \,,
\end{equation}
where $\delta^{2}\Psi=0$ denotes that we should keep the first order variation terms. This expression clearly shows us that our modified canonical energy is related to the direct generalization of the second order Einstein tensor. Our construction may be regarded as providing the generalization of the construction by the second order Einstein tensor in Ref.~\cite{Habisohn:1986} beyond Einstein gravity. Furthermore, our construction shows clearly the dependence of the modified canonical energy only on EOM and it makes the covariance of the expression manifest. It would be useful to deal with the odd-dimensional case with the gravitational Chern-Simons term. Besides, our expression of the canonical energy clarifies the relation between the traditional ADT expression and the canonical energy by Hollands and Wald.

One may worry that our form of the canonical energy might have some drawbacks compared to the HW construction since  it differs from the HW canonical energy  in the boundary term and the HW canonical energy is shown to be consistent with other criteria of the black hole stability.  Nevertheless,  as was shown in the above, all the essential properties of the canonical energy also hold in our modified version and there are various cases in which such a boundary term does not contribute.   In fact, our construction reveals the interesting aspect that the additional boundary term at the horizon may be allowed in  defining the canonical energy. Namely, when we define the canonical energy, we may relax its relation to the Hessian in thermodynamic stability, up to the boundary term. 
More concretely, we may have defined the modified canonical energy by adding the boundary term coming from ${\bf J}^{\mu}(K\,;\,\delta \Psi,\delta \Psi)$. For instance, we may have introduced the modified canonical energy by using the current expression  $ - {\cal J}^{\mu}_{ADT} +{\bf J}^{\mu} = {\cal J}^{\mu}_{\Delta}$ as
\begin{equation} \label{TilCan}  \nn 
\tilde{\cal E}(K\,;\, \delta \Psi, \delta \Psi) \equiv \frac{1}{8\pi G}\int_{\Sigma}dx_{\mu}\sqrt{-g}{\cal J}^{\mu}_{\Delta}(\Psi\,|,\Lie_{K}\delta \Psi,\delta \Psi)\,, 
\end{equation}
which also satisfies all the properties discussed in the above. This canonical energy can be shown to be different from the HW canonical energy as
\begin{align}   \label{} \nn
{\cal E}_{HW}(K\,;\, \delta\Psi,\delta\Psi)- \tilde{\cal E} (K\,;\, \delta\Psi,\delta\Psi)  =  \int_{B}dx_{\mu\nu}~   \sqrt{-g}{\bf A}^{\mu\nu}(\Psi\,|\,\Lie_{K}\delta\Psi, \delta\Psi)\Big|_{on-shell}~,
\end{align}
and satisfies the relation, when the second order EOM are imposed, as 
\begin{align}   \label{Hess}
\tilde{\cal E} (K_{H}\,;\, \delta\Psi,\delta\Psi)= \delta^{2}M_{\infty}- \Omega_{H}\delta^{2} J_{\infty} - \frac{\kappa}{2\pi}\delta^{2} \CS_{BH}\,,
\end{align}
which is consistent even with the Hessian in thermodynamic stability consideration. 

Recently, there was a suggestion that the canonical energy is dual to the so-called Fisher quantum information metric in the context of the AdS/CFT correspondence~\cite{Lashkari:2015hha}. As is clear from our construction, our modified canonical energy ${\cal E}$ or $\tilde{\cal E}$ is also a good candidate like those dual to the information metric, since our modified canonical energy does not give any difference from the HW canonical energy on the pure AdS background. The difference between them comes from the boundary contribution at the bifurcation surface $B$ or more correctly at $B(t)$, which is related to the  deep infrared physics in the boundary theory.  The  freedom of adding the boundary term $\int_{B}$ to  the canonical energy with arbitrary coefficients, may be useful in this duality.

In the following section, we consider hairy black holes in the asymptotic AdS space. In the asymptotic AdS space, the roles of boundary terms at infinity are irrelevant because of the AdS nature and one can take $K$ as the horizon Killing vector.  We apply our modified canonical energy to study the stability issue on hairy extremally rotating black holes.

\section{Hairy AdS black holes}

In this section we consider three-dimensional extremally rotating hairy AdS black holes admitted in Einstein gravity with a cosmological constant and a scalar field, whose analytic solutions are given in~\cite{Kwon:2012zh,Hyun:2012bc,Hotta:2008xt}. Interestingly, there are two arguments for the stability of the above extremally rotating hairy black holes that could give us opposite conclusions. The argument for their stability may be given as follows. Since there are no propagating degrees of freedom in three-dimensional Einstein gravity and the scalar field involved in the above solutions satisfies the Breitenlohner-Freedman bound~\cite{Breitenlohner:1982jf}, the extremal hairy black holes should be stable, at least, perturbatively.  Moreover, there seems to be no mechanism for the instability in this extremal configuration in the AdS/CFT context since it is dual to the renormalization group flow interpolating two CFTs, which does not seem to allow the other end points.  The opposite argument comes from the no-hair conjecture for AdS black holes~\cite{Hertog:2006rr,Hertog:2004bb}, which was made only for the four-dimensional case but seems to hold even in the three-dimensional case. Though there is a numerical attempt to construct rotating hairy black holes deformed from BTZ black holes~\cite{Banados:1992wn}, those hairy black holes require special conditions on the asymptotic behavior of the scalar field~\cite{Iizuka:2015vsa} that are not satisfied by the hairy extremal black holes under the consideration. Furthermore, the extremally rotating black holes in higher than four dimensions are shown to be unstable~\cite{Hollands:2014lra,Durkee:2010ea}. Of course, all the opposite arguments rely on the higher-dimensional analogues and so may not be so persuasive. In the following we adopt the canonical energy method and show the stability of three-dimensional extremally rotating hairy black holes.

Before presenting the specific models under  consideration, let us present some general setup for the Einstein gravity with  $U(1)$ gauge and scalar fields $\varphi^{I}$ and summarize some results.
  The Lagrangian for this system  consists of three parts, the Einsten-Hilbert one $\CL_{EH}$, the scalar one, and the $U(1)$ gauge part, respectively, as
\begin{equation} \label{} 
\CL_{EH}= R - 2\Lambda\,, \qquad 
\CL_{\varphi} = -\frac{1}{2}G_{IJ}\p_{\mu}\varphi^{I}\p^{\mu}\varphi^{J} - V(\varphi)\,, \qquad \CL_{A} = -\frac{1}{4}\CN(\varphi)F_{\mu\nu}F^{\mu\nu}\,.
\end{equation}
The Euler-Lagrange expressions for metric, gauge  and scalar fields are given by
\begin{align}   \label{}
\CE_{\mu\nu} &= \CG^{\Lambda}_{\mu\nu} - T_{\mu\nu}\,, \qquad \CE^{\mu}_{A} = \nabla_{\mu}(\CN F^{\mu\nu})\,, \nn \\
\CE_{\varphi} &= G_{IJ}(\varphi)(\Box \varphi^{J} +\Gamma^{J}_{KL}\p_{\mu}\varphi^{K}\p^{\mu}\varphi^{L}) - \p_{\varphi^{I}}V(\varphi) -\frac{1}{4}\p_{\varphi^{I}}\CN F_{\mu\nu}F^{\mu\nu}\,, \nn  
\end{align}
where $\CG^{\Lambda}_{\mu\nu} = R_{\mu\nu} - \frac{1}{2}R g_{\mu\nu}+\Lambda g_{\mu\nu}$ and the energy-momentum tensor $T_{\mu\nu}$ is composed of $T^{\varphi}_{\mu\nu}$ and $T^{A}_{\mu\nu}$ as
\[   
T^{\varphi}_{\mu\nu}= \frac{1}{2} \Big[G_{IJ}\p_{\mu}\varphi^{I}\p_{\nu}\varphi^{J} +g_{\mu\nu}\CL_{\varphi}\Big]\,, \qquad T_{\mu\nu}^{A} = \frac{1}{2} \Big[\CN F_{\mu\alpha}F_{\nu}\,^{\alpha} +g_{\mu\nu} \CL_{A}\Big]\,.
\]
Note that there is the off-shell identity for a generic diffeomorphism parameter $\zeta$ as 
\begin{align}   \label{}
0 &=-2T^{\mu\nu}_{A}\zeta_{\nu}+\zeta^{\mu}\CL_{A} - {\bf \Theta}^{\mu}_{A}(\Lie'_{\zeta}A)\,, \nn\\
0 &=-2T^{\mu\nu}_{\varphi}\zeta_{\nu} + \zeta^{\mu}\CL_{\varphi} - {\bf \Theta}^{\mu}(\Lie_{\zeta}\varphi)\,. \nn
\end{align}
where $\Lie'_{\zeta}A = -F_{\mu\nu}\zeta^{\nu}$ denotes the Lie derivative augmented by a gauge transformation and  the surface terms for the generic variations are given by 
\[
{\bf \Theta}^{\mu}_{g}(\delta g) = 2g^{\alpha[\mu} \nabla^{\beta]}\delta g_{\alpha\beta}\,,  \qquad 
 {\bf  \Theta}^{\mu} (\delta \varphi) = - G_{IJ}(\varphi) \delta \varphi^{I} \p^{\mu}\varphi^{J}\,, \qquad {\bf \Theta}^{\mu}_{A} = -\CN F^{\mu\nu}\delta A_{\nu}\,.
\]
By using this identity under the assumption $\delta K^{\mu}=0$, one can see that 
the scalar field and the Abelian gauge field parts for the modified canonical energy can be extracted from the surface term ${\bf \Theta}^{\mu}$ as
\begin{align}   \label{}
 \CJ^{\mu}_{ADT}(K\,;\, \delta\varphi, \delta\varphi)\Big|_{on-shell} &= \frac{1}{2} \Big[ - K^{\mu}\nabla_{\nu}\delta_{\varphi}  {\bf \Theta}^{\nu}_{\varphi}(\delta\varphi) +\delta_{\varphi}^2{\bf \Theta}^{\mu}(\pounds_{K}\varphi) \Big]
 \Big|_{on-shell}^{\delta^2 \varphi =0}\,, \nn \\
 \CJ^{\mu}_{ADT}(K\,;\, \delta A, \delta A)\Big|_{on-shell} &= \frac{1}{2}\Big[ - K^{\mu}\nabla_{\nu}\delta_{A}  {\bf \Theta}^{\nu}_{A}(\delta A) + \delta_{A}^{2}{\bf \Theta}^{\mu}(\pounds_{K}A) \Big]\Big|_{on-shell}^{\delta^2 A=0} \,. \nn \
\end{align} 
As was emphasized before, one can obtain the same expression solely from the expression of ${\bf E}^{\mu\nu}$, {\it i.e.} a combination of EOM, but we have provided the shortcut to the results by using the relation between EOM and the surface term ${\bf \Theta}^{\mu}$\,.


Now, let us stick to the three-dimensional Einstein gravity with a minimally  coupled scalar field, whose Lagrangian can be written as
\begin{equation} \label{}
\CL = R -\frac{1}{2}\p_{\mu}\varphi\p^{\mu}\varphi -V(\varphi)\,.
\end{equation}
Our interest is in the hairy deformed three-dimensional extremal black holes~\cite{Kwon:2012zh,Hyun:2012bc}. One can obtain the solutions by assuming that the scalar potential $V$ is taken in the form of 
\begin{equation} \label{}
V(\varphi) = \frac{1}{2L^{2}}(\p_{\varphi}\CW)^{2} -\frac{1}{2L^{2}}\CW^{2}\,, \qquad \CW = \CW(\varphi)\,.
\end{equation}
By taking the generic ansatz for the metric and scalar as
\begin{equation} \label{}
ds^{2} = -e^{2A(r)}dt^{2} + e^{2B(r)}dr^{2} +r^{2}(d\theta +e^{C(r)}dt)^{2}\,,\qquad \varphi =\varphi(r)\,,
\end{equation}
where the radius of the asymptotic $AdS_{3}$ space is taken to be unity, one can show that the metric functions and the scalar field satisfying the following first order ordinary differential equations solve the full EOM:
\begin{equation} \label{}
\varphi' = -e^{B}\p_\varphi\CW\,, \quad A' = \frac{1}{r} + e^{B}\CW\,, \quad (e^{C})' = \Big(\frac{1}{r}e^{A}\Big)'\,, \quad A' +B' = \frac{r}{2}\varphi'^{2}\,.
\end{equation}
For instance,  the simplest case among analytic solutions is 
\begin{equation} \label{}
\varphi (r) = \frac{\varphi_{0}}{r^{2}}\,, \qquad \CW = \alpha \Big[ 4 + \varphi^{2}(r)\Big] + \beta e^{\frac{\varphi^{2}}{4}}\,,
\end{equation}
where  the coefficients $\alpha$ and $\beta$ are given, in terms of the constant $\varphi_0$ and the position of the horizon $r_H$,  by
\begin{equation} \label{}
\alpha = \frac{1}{2}\frac{1}{1-e^{-\varphi^{2}_{0}/4r^{2}_{H}} } \,,  \qquad
\beta = - \frac{2e^{-\varphi^{2}_{0}/4r^{2}_{H}} }{1-e^{-\varphi^{2}_{0}/4r^{2}_{H}}}\,.
\end{equation}
In this case, the metric functions can be obtained as
\begin{equation} \label{}
e^{A} = r\Big[2 \alpha e^{-\varphi^{2}_{0}/4r^{2}}  + \frac{\beta}{2}\Big]\,, \qquad  e^{B} = e^{-\varphi^{2}_{0}/4r^{2}}  e^{-A}\,, \qquad e^{C} =  \frac{1}{r}e^{A}\,.
\end{equation}

The near horizon geometry of all these configurations satisfying the first order EOM are given generically by
\begin{equation} \label{}
ds^{2}_{NH} = L^{2}_{NH}\Big[-\rho^{2}dt^{2} + \frac{1}{\rho^{2}}d\rho^{2} \Big]+ r^{2}_{H}\Big(d\theta - \frac{L_{NH}}{r_{H}}\rho\, dt \Big)^{2}\,,
\end{equation}
which is known as the self-dual orbifold of $AdS_{3}$ space~\cite{Banados:1992gq}. Here, $L_{NH}$  denotes the radius of the orbifold of $AdS_{3}$ space defined by 
\[   
L_{NH} = \frac{1}{\CW(\varphi_{H})}\,.
\]
The scalar potential near the horizon can be expanded as
\begin{equation} \label{}
V = -\frac{1}{2}\CW(\varphi_{H})^{2} + \CW(\varphi_{H})^{2}(\varphi -\varphi_{H})^{2} + \cdots\,,
\end{equation}
which comes from the generic expansion of the superpotential ${\cal W}$ as 
\begin{equation} \label{}
\CW = \CW(\varphi_{H}) - \frac{1}{2}\CW(\varphi_{H})(\varphi-\varphi_{H})^{2} + \cdots\,.
\end{equation}
Note that the first term of the superpotential $\CW$ plays the role of the cosmological constant on the near horizon geometry. 
The horizon Killing vector is taken by $K=\frac{\partial}{\partial t}$ in these coordinates.
The effective Lagrangian on the near horizon geometry is given by 
\begin{equation} \label{}
\CL = R- 2\Lambda_{NH}-\frac{1}{2}(\p_{\mu}\tilde\varphi)^{2} - \frac{1}{2} m^{2}_{NH}( \tilde\varphi)^{2} +\cdots\,,
\end{equation}
where $\Lambda_{NH}$ and $m^{2}_{NH}$ denote the near horizon effective cosmological constant and effective mass square of the scalar field $ \tilde\varphi \equiv \varphi - \varphi_{H}$ as 
\[   
\Lambda_{NH}\equiv -\frac{1}{4L^{2}_{NH}} = - \frac{\CW(\varphi_H)^{2}}{4} \,, \qquad m^{2}_{NH} = 2\CW(\varphi_H)^{2}= \frac{2}{L^{2}_{NH}}\,.
\]
The effective mass of the scalar field  is greater than the Breitenlohner-Freedman bound in the near horizon geometry and so the scalar field could be thought of as  stable in the near horizon geometry. We would like to confirm this argument explicitly by using the canonical energy method.

Let us consider linear perturbations of the metric and scalar fields and compute the modified canonical energy  on the near horizon geometry in order to see the stability  of the above hairy deformed extremal black holes. Combined with the stability argument at infinity, one may say that the whole configuration is stable for linear perturbations. As is obvious from the three-dimensional nature of our configurations, the metric variation should be just pure gauge and so its role is trivial. From now on, let us take all the functions to depend on the radial coordinate $\rho$ instead of $r$ on the near horizon geometry. 
 Indeed, by taking the metric perturbation as 
\begin{equation} \label{}
\delta g_{\m\n} = \nabla_{(\mu}\zeta_{\nu)}\,,\qquad \zeta^\mu=\zeta^\mu(t,\rho,\theta)\,,
\end{equation}
we will show that the metric perturbation does not contribute to the canonical energy. 
%

Instead of solving the linearized EOM on the given background, we would like to analyze the form of modified canonical energy itself, which is composed of four parts as 
\begin{align} \label{}
\CE(K\,;\, \delta_{1} \Psi,\delta_{2}\Psi) = \CE(K\,;\, \delta_{1} g,\delta_{2}g) + \CE(K\,;\, \delta_{1} g,\delta_{2}\varphi) + \CE(K\,;\, \delta_{1} \varphi,\delta_{2}g) + \CE(K\,;\, \delta_{1} \varphi,\delta_{2}\varphi)\,.
\end{align}
By using our result given in Eq.~(\ref{modCE}), one can compute each term directly without difficulty. Firstly, the contribution  from the metric perturbation to the canonical energy is given by\footnote{It would be meaningful to check the gauge invariance explicitly since the gauge choice may be different in the extremal case and in our form of the near horizon geometry.}
\begin{align}
&\CE(K\,;\, \delta g,\delta g) \\
&=  -\frac{1}{8\pi G}\int_{\Sigma}dx_{\mu}~ \sqrt{-g}\bigg[ \Big[ -h^{\r\s}\nabla^{\m}\nabla^{\n}h_{\r\s} + 2 h^{\r\s}\nabla_{\r}\nabla^{(\m}h^{\n)}\,_{\s} -\frac{1}{2}\nabla^{\m}h^{\r\s}\nabla^{\n}h_{\r\s} -2\nabla^{\r}h^{\s\m}\nabla_{[\r}h_{\s]}\,^{\n} \nn\\
&\quad -\nabla_{\r}(h^{\r\s}\nabla_{\s}h^{\m\n}) +\frac{1}{2}\nabla^{\r}h\nabla_{\r}h^{\m\n} +2\Big(\nabla_{\r}h^{\r\s} -\frac{1}{2}\nabla^{\s}h\Big)\nabla^{(\m}h^{\n)}\,_{\s} \Big]   -\frac{1}{2}g^{\m\n} \Big[ trace \Big] \bigg]K_{\nu}\Big|^{\delta^{2}\Psi=0}_{on-shell}\,, \nn
\end{align}
where $h_{\m\n}\equiv\d g_{\m\n}$, $h\equiv g^{\m\n}h_{\m\n}$ and $[trace]$ denotes the trace of the expression in front of it. This part is consistent with Eq.~(85) in \cite{Hollands:2012sf}. Since the metric perturbation is given by pure gauge transformation, {\it i.e.} $\delta g_{\mu\nu}=\pounds_\zeta g_{\mu\nu}$, this part has nothing to do with canonical energy and can be checked to vanish by a direct computation, as was shown generically in Eq.~(\ref{CanEgauge2}). The cross terms can be shown to vanish as follows:
\begin{align} \label{}
&\CE(K\,;\, \delta g,\delta\varphi) + \CE(K\,;\, \delta \varphi,\delta g) \\
&= \frac{1}{8\pi G}\int_{\Sigma}dx_{\mu}~ \sqrt{-g} \Big[ m_H^2(2h_{\mu\nu} - g_{\mu\nu} h)\varphi \delta\varphi \Big] K_{\nu}\Big|^{\delta^{2}\Psi=0}_{on-shell} =0 \,.\nn
\end{align}
Hence, the  contribution to canonical energy comes only from the scalar perturbation part as 
\begin{align} \label{}
\CE(K\,;\, \delta \varphi,\delta \varphi) &= -\frac{1}{8\pi G}\int_{\Sigma}dx_{\mu}~ \sqrt{-g} \Big[ -\nabla^\mu\delta\varphi\nabla^\nu\delta\varphi+\frac{1}{2}g^{\mu\nu}\Big( \nabla_\lambda\delta\varphi\nabla^\lambda\delta\varphi + m_H^2 \delta\varphi^2 \Big) \Big] K_{\nu}\Big|^{\delta^{2}\Psi=0}_{on-shell} \nn\\
&= \frac{1}{8\pi G}\int_{\Sigma}d\rho d\theta\, \frac{r_{NH}}{2\rho^2} \bigg[ 2\rho^2 \d\f^2 + \rho^4 \Big( \frac{\partial \d\f}{\partial \rho} \Big)^2 + \Big( \frac{\partial \d\f}{\partial t} \Big)^2 \bigg]\,.
\end{align}
Since $r_{NH}>0$, $\CE(K\,;\, \delta \varphi,\delta \varphi)$ could not be negative at any time. This confirms the linear stability of the extremally rotating hairy black holes under consideration.

\section{Conclusion}

We have constructed the modified version of the canonical energy that was introduced originally by HW in~\cite{Hollands:2012sf}. Our construction is based on the off-shell adaptation of the ADT current and so connects the various conceptually different constructions. Briefly speaking, it can be regarded as the generalization of the second order Einstein tensor method in pure Einstein gravity or the effective energy-momentum tensor method in the original ADT approach.
By showing explicitly  the relation between our construction and the original HW one, we have showed that one may construct a quantity that differs from the HW canonical energy in the boundary term over the spatial section of the future horizon. Through this relation, we have also explained clearly why the second order Einstein tensor method in the literature could give the same information as the HW canonical energy at the asymptotic infinity. In other words, our results imply that the second order contribution to the Bondi energy can be computed by using the ADT current expression.

In fact, the modified canonical energy can be constructed while sharing all the properties of the HW canonical energy as given in Eq.~(\ref{TilCan}) and may be distinguished from the HW canonical energy only in the higher derivative theory of gravity.  The essential point of  our construction is that one may have  freedom in constructing the {\it canonical energy} equipped with the relevant properties.  This  possibility would give us a better chance to  match the canonical energy to the Fisher information metric in the context of the AdS/CFT correspondence. Our results show that one may be able to use the  freedom in the construction of  the canonical energy  with the required properties under consideration. 

We have also considered the three-dimensional extremally rotating hairy AdS black holes that were not yet proven to  be stable or not. Since there are conflicting arguments about their stability, it would be a good exercise to use the (modified) canonical energy method in this example, as is done in the main text. We have verified that the canonical energy is positive definite on the near horizon geometry and concluded that the extremally rotating hairy black holes in three dimensions are stable at least under the linear perturbations.   It would be very interesting to explore whether or not one can distinguish the various possible forms of the canonical energy from the physical consideration  in the context of the AdS/CFT correspondence, especially as a dual to the Fisher information metric. In the context of the black hole stability, it would also be interesting to consider a higher derivative theory of gravity and the stability of its black hole solutions by using the (modified) canonical energy method.

\vskip 1cm
\centerline{\large \bf Acknowledgments}
\vskip0.5cm
{We would like to thank  Nakwoo Kim for some discussion.
SH was supported by the National Research Foundation of Korea(NRF) grant 
with the grant number NRF-2013R1A1A2011548. SY was supported by the National Research Foundation of Korea(NRF) grant with the grant number NRF-2015R1D1A1A09057057.}
%

{\center \section*{Appendix}}

\section*{A. Relation to symplectic current}
\renewcommand{\theequation}{A.\arabic{equation}}
  \setcounter{equation}{0}

Generic variation of the action can be expressed as 
\begin{equation} \label{}
\delta I[\Psi] =\frac{1}{16\pi G}\int d^{D}x ~\delta (\sqrt{-g}\CL) =  \frac{1}{16\pi G}\int d^{D}x \Big[\sqrt{-g}\CE_{\Psi} \delta \Psi + \p_{\mu}\Theta^{\mu}(\Psi,\delta \Psi)\Big]\,.
\end{equation}
Identifying the diffeomorphism transformation of the parameter $\zeta$ with the above generic variation leads to the following relation: 
\begin{equation} \label{Diffeo}
\p_{\mu}(\zeta^{\mu}\sqrt{-g}\CL) = \sqrt{-g}\CE_{\Psi}\Lie_{\zeta}\Psi + \p_{\mu}\Theta^{\mu}(\Psi\,|\,\Lie_{ \zeta}\Psi)\,.
\end{equation}
The symplectic current with the variation of the diffeomorphism parameter $\zeta^{\mu}$ under a generic variation $\delta$ can be defined by
\begin{align}   \label{}
\omega^{\mu} (\Psi\,|\, \Lie_{\zeta}\Psi, \delta \Psi) &\equiv \Lie_{\zeta}\Theta^{\mu}(\Psi,\delta \Psi) -  \Big[\delta\{\Theta^{\mu}(\Psi,\Lie_{\zeta}\Psi)\} - \Theta^{\mu}(\Psi,\Lie_{\delta \zeta}\Psi)\Big]  \nn \\
&= \Theta^{\mu}(\Lie_{\zeta}\Psi,\delta \Psi) - \Theta^{\mu}(\delta\Psi,\Lie_{\zeta} \Psi)\,,  \nn
\end{align}
which reduces to the conventional one when $\delta \zeta^{\mu}=0$. By combining  the definition of the symplectic current $\omega^{\mu}(\Psi\,|\, \Lie_{\zeta}\Psi, \delta\Psi)$, the double variation of the action as $( \delta\Lie_{\zeta} - \Lie_{\zeta}\delta)I[\Psi]= \Lie_{\delta \zeta}I [\Psi]$, and the relation  in~(\ref{Diffeo}) for the diffeomorphism parameter $\delta \zeta^{\mu}$, one obtains  
\begin{equation} \label{identity}
\p_{\mu}\omega^{\mu}(\Psi\,|\, \Lie_{\zeta}\Psi, \delta\Psi) = \Big[ \delta\Big(\sqrt{-g}\CE_{\Psi}\Lie_{\zeta}\Psi \Big)  - \Big( \sqrt{-g}\CE_{\Psi}\Lie_{\delta \zeta}\Psi\Big) \Big] - \Lie_{\zeta} \Big(\sqrt{-g}\CE_{\Psi}\delta\Psi \Big)\,.
\end{equation}
By using  the identity in Eq.~(\ref{genBI}),  the relation of $\delta\p_{\mu}=\p_{\mu}\delta$ for a generic variation $\delta$, and the property of the Lie derivative on the scalar density $\Lie_{\zeta}(\sqrt{-g}\CE_{\Psi}\delta \Psi) = \p_{\mu} (\zeta^{\mu}\sqrt{-g}\CE_{\Psi}\delta \Psi)$, one can see that 
\begin{equation} \label{conA}
\p_{\mu}\Big(\sqrt{-g}\CJ^{\mu}_{ADT}\Big) = -\frac{1}{2}\Big[\delta \Big(\sqrt{-g}\CE_{\Psi}\Lie_{\zeta}\Psi \Big) -  \sqrt{-g}\CE_{\Psi}\Lie_{\delta \zeta}\Psi\Big]  + \frac{1}{2}\Lie_{\zeta} \Big(\sqrt{-g}\CE_{\Psi}\delta\Psi \Big)\,.
\end{equation}

The additional current  ${\cal J}^{\mu}_{\Delta}$ can  be related to  the symplectic current $\omega^{\mu} $ in the covariant phase space~\cite{Lee:1990nz}
as
\begin{equation} \label{AddCur}
2\sqrt{-g} \CJ^{\mu}_{\Delta} (\Psi\,|\,\Lie_{\zeta} \Psi, \delta \Psi) = \omega^{\mu} (\Psi\,|\,\Lie_{\zeta} \Psi, \delta \Psi) + \p_{\nu}\Big(\sqrt{-g}{\bf A}^{\mu\nu}  (\Psi\,|\,\Lie_{\zeta} \Psi, \delta \Psi) \Big)\,,
\end{equation}
where ${\bf A}^{\mu\nu}$ is an anti-symmetric tensor defined by
\[   
\delta \Theta^{\mu}(\Lie_{\zeta} \Psi) = \Lie_{\zeta}\Theta^{\mu}(\delta \Psi) + \sqrt{-g}\nabla_{\nu}\Big({\bf A}^{\mu\nu}(\Psi\,|\,\Lie_{\zeta} \Psi, \delta \Psi) - 2{\bf S}^{\mu\nu}(\Psi\,|\,\Lie_{\zeta} \Psi, \delta \Psi)\Big)+ \delta \Psi\, [ \cdots ]\,,  
\]
where ${\bf S}^{\mu\nu} \equiv {\bf S}^{(\mu\nu)}$ and $[\cdots]$ denotes the irrelevant expressions in our presentation. As a result,  the additional current ${\cal J}^{\mu}_{\Delta}$ is symplectic just as $\omega^{\mu}$ and  vanishes for a Killing vector. The relation in Eq.~(\ref{AddCur}) between the additional current term $\CJ^{\mu}_{\Delta}$ and the symplectic current $\omega^{\mu}$  implies that
\begin{equation} \label{}
\p_{\mu}(\sqrt{-g}\CJ^{\mu}_{\Delta}) =\frac{1}{2}\p_{\mu}\omega^{\mu}\,.
\end{equation}
Now, the identical conservation of the current ${\bf J}^{\mu}$ follows from the identity given in Eq.~(\ref{identity}) 
\begin{align}   \label{}
\p_{\mu}(\sqrt{-g}{\bf J}^{\mu}) &= \p_{\mu}(\sqrt{-g}\CJ^{\mu}_{ADT}) + \p_{\mu}(\sqrt{-g}\CJ^{\mu}_{\Delta}) \nn \\
&= -\frac{1}{2}\Big[\delta \Big(\sqrt{-g}\CE_{\Psi}\Lie_{\zeta}\Psi \Big) -  \sqrt{-g}\CE_{\Psi}\Lie_{\delta \zeta}\Psi\Big] + \frac{1}{2}\Lie_{\zeta} \Big(\sqrt{-g}\CE_{\Psi}\delta\Psi \Big)  + \frac{1}{2}\p_{\mu}\omega^{\mu}(\Psi\,|\, \Lie_{\zeta}\Psi, \delta \Psi) \nn \\
&=0\,. \nn
\end{align}

For a covariant Lagrangian $L(\Psi)$,
 the off-shell Noether current and potential may be introduced as
\begin{equation} \label{}
J^{\mu} =\zeta^{\mu}\sqrt{-g}L(\Psi) +  2\sqrt{-g}{\bf E}^{\mu\nu}\zeta_{\nu} - \Theta^{\mu}(\Lie_{\zeta} \Psi)=\p_{\nu}K^{\mu\nu}(\zeta)\,.
\end{equation}
After some manipulation by using the relation in Eq.~(\ref{AddCur}), one can obtain the off-shell relation 
\begin{equation} \label{ADTpot}
2\sqrt{-g}{\bf Q}^{\mu\nu}(\zeta\,;\,\Psi,\delta \Psi) = \delta K^{\mu\nu}(\zeta)  - K^{\mu\nu}(\delta \zeta) -2\zeta^{[\mu}\Theta^{\nu]}(\delta \Psi)+ \sqrt{-g}{\bf A}^{\mu\nu}(\Psi\,|\,\Lie_{\zeta}\Psi,\delta \Psi)\,.
\end{equation}
Note that there may be  additional terms in the right-hand side in the above relation when the Lagrangian contains non-covariant terms~\cite{Kim:2013cor}.  By recalling  the relations Eqs.~(\ref{DefOffCur}), (\ref{AddCur}), and (\ref{ADTcur}),  one may note   that
\begin{align}   \label{RelsymCurexp}
   \omega^{\mu}(\Psi\,|\, \Lie_{\zeta} \Psi,\delta\Psi) + 2\sqrt{-g}{\cal J}^{\mu}_{ADT}(\zeta\,;\,\Psi,\delta\Psi)   & =  \partial_{\nu} \Big[2\sqrt{-g}{\bf Q}^{\mu\nu}(\zeta\,;\,\Psi, \delta\Psi)  - \sqrt{-g}{\bf A}^{\mu\nu}(\Psi\,|\,\Lie_{\zeta}\Psi, \delta\Psi)\Big]  \nn \\ 
 &= \partial_{\nu} \Big[  \delta  K^{\mu\nu}(\zeta)  - K^{\mu\nu}(\delta \zeta) -2\zeta^{[\mu}\Theta^{\nu]}(\delta \Psi) \Big]\,, 
\end{align}
where we used the off-shell identity~(\ref{ADTpot}) in the second equality. 
%

It is straightforward to repeat the same procedure in~\cite{Wald:1993nt} to derive the first law of black hole thermodynamics. 
Let us introduce the integral $\CV$ of the off-shell current ${\bf J}^{\mu}$ as\footnote{Since we use a Killing vector in the derivation of the first law, one can  employ $\CJ^{\mu}_{ADT}$ instead of ${\bf J}^{\mu}$.}
\begin{align}   \label{}
\CV_{\Sigma} (\zeta\,;\, \Psi,\delta \Psi) & \equiv \frac{1}{8\pi G}\int_{\Sigma}dx_{\mu}\sqrt{-g}{\bf J}^{\mu}(\zeta\,;\, \Psi,\delta \Psi)   \nn \\
&=\frac{1}{8\pi G} \int_{\infty}dx_{\mu\nu}\sqrt{-g}{\bf Q}^{\mu\nu} - \frac{1}{8\pi G} \int_{B}dx_{\mu\nu}\sqrt{-g}{\bf Q}^{\mu\nu}\,.
\end{align}
To see the implication of this integral, take the on-shell condition and $\zeta$  as a Killing vector $K$. Then, the off-shell current ${\bf J}^{\mu}_{ADT}$ reduces to the ADT current ${\cal J}^{\mu}_{ADT}$. The on-shell condition implies $\CJ^{\mu}_{ADT}=0$ and the Killing condition leads to $\CJ^{\mu}_{\Delta}=0$ and ${\bf A}^{\mu\nu}=0$, and so the integral $\CV_{\Sigma}$ vanishes in this case.  For the horizon Killing vector $K_{H}$, 
\[   
K_{H} \equiv \frac{\p}{\p t} + \Omega_{H} \frac{\p}{\p \theta}\,,
\]
the conserved charges are given by
\begin{align}   \label{}
\frac{1}{8\pi G} \int_{\infty}dx_{\mu\nu}\sqrt{-g}{\bf Q}^{\mu\nu} (K_{H}\,;\, \Psi,\delta \Psi)
&= \delta M_{\infty}-\Omega_{H}\delta J_{\infty} \nn \\ 
\frac{1}{8\pi G} \int_{B}dx_{\mu\nu}\sqrt{-g}{\bf Q}^{\mu\nu} (K_{H}\,;\, \Psi,\delta \Psi)
 &=\frac{\kappa}{2\pi} \delta \CS_{BH}\,, \nn
\end{align}
and so
one can see that the integral gives us the first law of black hole thermodynamics as
\begin{equation} \label{}
0= \delta M_{\infty} - \Omega_{H}\delta J_{\infty} - \frac{\kappa}{2\pi}\delta \CS_{BH}\,.
\end{equation}
%

\section*{B. Gauge invariance}
\renewcommand{\theequation}{B.\arabic{equation}}
  \setcounter{equation}{0}
 
Just like two expressions of the Lie derivative of the ADT current in Eq.~(\ref{Gauge1}) and~(\ref{Gauge2}), the Lie derivative of the potential ${\bf Q}^{\mu\nu}$ can be written in two ways. Firstly, it can be written as
\begin{equation} \label{}
\Lie_{\epsilon}{\bf Q}^{\mu\nu}(K\,;\,\Psi, \delta \Psi) = {\bf Q}^{\mu\nu}(\Lie_{\epsilon}K\,;\,\Psi, \delta \Psi) + {\bf Q}^{\mu\nu}(K\,;\, \Lie_{\epsilon}\Psi, \delta \Psi) +{\bf Q}^{\mu\nu}(K\,;\,\Psi, \Lie_{\epsilon}\delta \Psi)\,.
\end{equation}
Secondly, it can  also be  written as
\begin{align}   \label{}
\Lie_{\epsilon}{\bf Q}^{\mu\nu}(K\,;\,\Psi, \delta \Psi) &= \nabla_{\alpha}(3\epsilon^{[\alpha}{\bf Q}^{\mu\nu]}) -{\bf Q}^{\mu\nu}\nabla_{\alpha}\epsilon^{\alpha} - \epsilon^{\mu}\nabla_{\alpha}{\bf Q}^{\nu \alpha} - \epsilon^{\nu}\nabla_{\alpha}{\bf Q}^{\alpha\mu}\,, \nn \\
&= \nabla_{\alpha}(3\epsilon^{[\alpha}{\bf Q}^{\mu\nu]}) -{\bf Q}^{\mu\nu}\nabla_{\alpha}\epsilon^{\alpha} - 2\epsilon^{[\mu}{\cal J}^{\nu]}_{~ADT}\,,
\end{align}
where we have used  that ${\cal J}^{\mu}_{ADT} (K)= {\bf J}^{\mu}(K)=\nabla_{\nu}{\bf Q}^{\mu\nu}(K)$ for a Killing vector $K$.
Combining the above two expressions for the Lie derivative of the potential ${\bf Q}^{\mu\nu}$, one obtains
\begin{align}   \label{GaugeRel}
{\bf Q}^{\mu\nu}(K\,;\,\Psi, \Lie_{\epsilon}\delta \Psi)\Big|_{on-shell} &= \Big[\nabla_{\alpha}\Big(3\epsilon^{[\alpha}{\bf Q}^{\mu\nu]}(K\,;\,\Psi, \Lie_{\epsilon}\delta \Psi)\Big)   -{\bf Q}^{\mu\nu}(K\,;\,\Psi, \delta \Psi)\nabla_{\alpha}\epsilon^{\alpha} \nn \\
&\qquad   -  {\bf Q}^{\mu\nu}(K\,;\, \Lie_{\epsilon}\Psi, \delta \Psi) - {\bf Q}^{\mu\nu}(\Lie_{\epsilon}K\,;\,\Psi, \delta \Psi) \Big]_{on-shell} \,,
\end{align}
where the ``on-shell'' vanishing condition of the ADT current ${\cal J}^{\mu}_{ADT}$ is used. By inserting  this equality in Eq.~(\ref{CanEgauge}),  one can see that there are contributions from two boundaries: the spacelike infinity and the bifurcation  surface $B$. The first term in the right-hand side of the above equality does not contribute to the modified canonical energy since we integrate over the closed space at both boundaries.  Now, let us consider the leftover terms in the right-hand side. 

Since $\epsilon$ corresponds to the asymptotic symmetry generators at  infinity, one can see that $\nabla_{\alpha}\epsilon^{\alpha}\rightarrow 0$  and $\Lie_{\epsilon}\Psi\rightarrow 0$ sufficiently fast near infinity compared to the field $\Psi$ itself, which would come from the definition of the asymptotic symmetry generators. As can be inferred from the definition of the charge in Eq.~(\ref{ADTcharge}),  it turns out that $\delta Q(K) \simeq\int dx_{\mu\nu}\sqrt{-g} {\bf Q}^{\mu\nu}(K;\, \Psi, \delta \Psi)$ is finite (in fact, taken as zero for linear perturbations) at the spacelike infinity. Therefore, the second and third terms in the right-hand side vanish at the spacelike infinity. Furthermore, $\Lie_{\epsilon}K = [ \epsilon, K] =\epsilon'$ corresponds to another asymptotic Killing vector and  $\delta Q(\epsilon') \simeq \int dx_{\mu\nu} \sqrt{-g}{\bf Q}^{\mu\nu}(\epsilon'\,;\, \Psi, \delta \Psi) =0$ under the chosen condition that the charge is invariant for the linear perturbations. As a result, the last term contributes at the spacelike infinity. 

 Since we are taking the same gauge conditions in Ref.~\cite{Hollands:2012sf}, our gauge parameter $\epsilon$ satisfies the same property as there. Thus, at the bifurcation surface $B$, the gauge transformation satisfies $\nabla_{\alpha}\epsilon^{\alpha} |_{B} =\mu^{\alpha\beta}\nabla_{\alpha}\epsilon_{\beta} |_{B}=0$  (see Remark below Lemma 1 in Ref.~\cite{Hollands:2012sf}). Therefore, the second term in the right-hand side in Eq.~(\ref{GaugeRel}) does not contribute.   At the bifurcation surface $B$, therefore, the  relevant expression  in Eq.~(\ref{GaugeRel}) can be written as
\begin{align}   \label{GaugeRel2}
2\sqrt{-g}{\bf Q}^{\mu\nu}(K\,;\,\Psi, \Lie_{\epsilon}\delta \Psi)\Big|_{on-shell} = -2\sqrt{-g}\Big[ {\bf Q}^{\mu\nu}(K\,;\, \Lie_{\epsilon}\Psi, \delta \Psi) + {\bf Q}^{\mu\nu}(\Lie_{\epsilon}K\,;\,\Psi, \delta \Psi)     \Big]_{on-shell} \,.   
\end{align}
For simplicity, let us focus on  pure Einstein gravity, in which the potential ${\bf Q}^{\mu\nu}$  is given by~\cite{Hyun:2014kfa}
\begin{equation} \label{}
{\bf Q}^{\mu\nu} (\zeta\,;\, g, h ) =  \frac{1}{2}h \nabla^{[\mu}\zeta^{\nu]}    -\zeta^{[\mu} \nabla_{\alpha} h^{\nu]\alpha} + \zeta_{\alpha}\nabla^{[\mu}h^{\nu]\alpha}  + \zeta^{[\mu}\nabla^{\nu]} h   - \frac{1}{2} h^{\alpha [\mu} \nabla_{\alpha} \zeta^{\nu]} + \frac{1}{2}h^{\alpha [\mu} \nabla^{\nu]} \zeta_{\alpha}\,,   
\end{equation}
By using  the metric perturbation near the horizon  given  in~Eq.~(\ref{PertH}), one can  see that the perturbation metric $h_{\mu\nu}$ satisfies $n_{[\mu}\ell_{\nu]}h^{\mu\alpha} \rightarrow 0$ as $r\rightarrow 0$, that is to say, $n_{[\mu}\ell_{\nu]}h^{\mu\alpha} $ vanishes on the future horizon. Therefore, the last two terms in the above potential ${\bf Q}^{\mu\nu}(\zeta\,;\, g, \delta g ) $ do not contribute after the integration over the spatial section $B(t)$. This is the case even for the second variation in the form of ${\bf Q}^{\mu\nu} (\zeta\,;\, \delta_{2} g, \delta_{1} g )$.

Let us consider the contribution from the first term  ${\bf Q}^{\mu\nu}(K\,;\, \Lie_{\epsilon}g, \delta g )$ in the right-hand side of Eq.~(\ref{GaugeRel2}).
By incorporating $K_{H} \rightarrow 0$  at the bifurcation surface $B$, this term  reduces to 
\begin{equation} \label{}
 {\bf Q}^{\mu\nu} (K_H\,;\, \Lie_{\epsilon} g, h)\Big|_{B} =  -\frac{1}{2} h^{\alpha\beta}\Lie_{\epsilon} g_{\alpha\beta}\nabla^{[\mu}K_H^{\nu]} \Big|_{B}\,.
\end{equation}
%
Note that the gauge parameter $\epsilon^\mu$ is tangent to the future horizon and, in fact, its admissible form is given by~\cite{Hollands:2014lra, Green:2015kur}:
\begin{equation} \label{}
\epsilon^{\mu} = f_{\epsilon}n^{\mu} + rY^{\mu}_{\epsilon}\,,\qquad n^\mu \nabla_\mu f_\epsilon =0\,.
\end{equation}
Then,   the direct computation in the chosen coordinates   in~Eq.~(\ref{PertH}) shows us that 
$h^{\mu\nu} \nabla_{[\mu}\epsilon_{\nu]} \rightarrow 0$ as $r\rightarrow0$. Hence, the first term gives zero contribution. 
  In the chosen gauge near the horizon, $ \xi\equiv  \Lie_{\epsilon}K=-[K, \epsilon]$  becomes normal to the future horizon at the surface $B$, since the gauge parameter $\epsilon^{\mu}$ is  tangent to the horizon. By using the property of $\xi$, one can set~\cite{Hollands:2012sf}
\begin{equation} \label{}
\xi^{\mu} = f n^{\mu} + u X^{\mu} + r Y^{\mu}\,.
\end{equation}
Noting that $n^{\mu}\nabla_{\alpha}h_{\mu}^{~\alpha}\rightarrow 0$ as $r\rightarrow 0$, 
with   the expression  of  $\xi$ near the horizon,  one can show that  the second term ${\bf Q}^{\mu\nu}(\Lie_{\epsilon}K\,;\,\Psi, \delta \Psi)$ in the right-hand side of Eq.~(\ref{GaugeRel2})  reduces at the surface $B$ to 
\begin{equation} \label{}
{\bf Q}^{\mu\nu} (\xi \,;\, g, \delta g) =    \frac{1}{2}\delta \mu^{\alpha}_{\alpha} \nabla^{[\mu}\xi^{\nu]} + \xi_{\alpha}\nabla^{[\mu}h^{\nu]\alpha}  + \xi^{[\mu}\nabla^{\nu]} h\,.
\end{equation}
Since  $h^{\mu\nu}\xi_{\nu} =h^{\mu\nu}n_{\nu}=0$ at the surface $B$, one can see that  $n_{[\mu}\ell_{\nu]}\xi_{\alpha}\nabla^{[\mu}h^{\nu]\alpha}=0$ for the metric perturbation $h_{\mu\nu}$ at $B$ and that $n^{\mu}\nabla_{\mu}h \propto \mu^{ab}\p_{u}\delta \mu_{ab} \propto \delta \vartheta =0$.
Thus, we obtain the following result:
\begin{equation} \label{}
2\sqrt{-g}{\bf Q}^{\mu\nu}(K\,;\,g, \Lie_{\epsilon}\delta g)\Big|_{on-shell} = -\sqrt{-g}\delta \mu^{\alpha}_{\alpha} \nabla^{[\mu}\xi^{\nu]}\,.
\end{equation}
Since the gauge is chosen as $\mu^{\alpha\beta}\delta \mu_{\alpha\beta}= \delta \mu^{\alpha}_{\alpha}=0$, 
we immediately see that
\begin{equation} \label{}
\CE(K\,;\, \Lie_{\epsilon} g,\delta g)  = \frac{1}{16\pi G}\int_{B} dx_{\mu\nu}\sqrt{-g}\delta \mu^{\alpha}_{\alpha} \nabla^{[\mu}\xi^{\nu]}=0\,.
\end{equation}

Now, we would like to give comments on the relation to the derivation in Ref.~\cite{Hollands:2012sf}. In short, our derivation is completely parallel and consistent to the one in Appendix A of Ref.~\cite{Hollands:2012sf}.
In fact, one can show that
\begin{align}
&2\sqrt{-g}{\bf Q}^{\mu\nu}(K\,;\,\Psi, \Lie_{\epsilon}\delta \Psi)\Big|_{on-shell} \nn \\
&= -\Big[  2 \sqrt{-g} {\bf Q}^{\mu\nu}(K\,;\, \Lie_{\epsilon}\Psi, \delta \Psi)  -\delta K^{\mu\nu}(\xi) + 2\xi^{[\mu}\Theta^{\nu]}(\delta \Psi)  -\sqrt{-g}  {\bf A}^{\mu\nu}(\Psi\,|\,\Lie_{\xi}\Psi, \delta\Psi) \Big]_{on-shell}\,.  \nn 
\end{align}
In pure Einstein gravity, the first term is already shown to give no contribution.
Note that the ${\bf A}^{\mu\nu}$-tensor is given  by
\begin{align}   \label{}
{\bf A}^{\mu\nu}(\Lie_{\xi}g, \delta g) &=   -\Big(g^{\mu(\alpha}g^{\beta)(\rho} g^{\sigma)\nu} - g^{\nu(\alpha}g^{\beta)(\rho}g^{\sigma)\mu} \Big)  (\Lie_{\xi}g_{\alpha\beta}h_{\rho\sigma}  - h_{\alpha\beta}\Lie_{\xi} g_{\rho\sigma} )\,,
\end{align}
from which one can see that  the above ${\bf A}^{\mu\nu}$-tensor term does not contribute to the canonical energy through Eqs.~(\ref{GaugeRel}),~(\ref{GaugeRel2}), and~(\ref{CanEgauge}).  The absence of the contribution from $\delta K^{\mu\nu}(\xi) - 2\xi^{[\mu}\Theta^{\nu]}(\delta \Psi)$ is the main result in Appendix A in~\cite{Hollands:2012sf}.

\section*{C. Relation to HW construction } 
\renewcommand{\theequation}{C.\arabic{equation}}
  \setcounter{equation}{0}

 In the case of the Killing vector $K$ with $\delta K=0$,   one can show that the current expression ${\cal J}^{\mu}_{ADT}(K\,;\, \delta_{2}\Psi,\delta_{1}\Psi)$ is related to the symplectic current as follows. The variation of the ADT current ${\cal J}^{\mu}_{ADT}$ can also be written  under the condition $\delta K=0$  as
\begin{align}   \label{}
\delta_{2}\Big(\sqrt{-g}{\cal J}^{\mu}_{ADT}(K\,;\,\Psi,\delta_{1}\Psi)\Big) &= \sqrt{-g} {\cal J}^{\mu}_{ADT}(K\,;\, \delta_{2}\Psi, \delta_{1}\Psi) + \sqrt{-g} {\cal J}^{\mu}_{ADT}(K\,;\, \Psi, \delta_{2}\delta_{1}\Psi)\nn \\
&=  \sqrt{-g} {\cal J}^{\mu}_{ADT}(K\,;\, \delta_{2}\Psi, \delta_{1}\Psi)  + \partial_{\nu}\Big[\sqrt{-g}{\bf Q}^{\mu\nu}(K\,;\, \Psi, \delta_{2}\delta_{1}\Psi)\Big]\,, \nn 
\end{align}
where we used  in the first equality  the ``on-shell'' vanishing condition of ${\cal J}^{\mu}_{ADT}(\zeta\,;\, \Psi, \delta \Psi)$ and used   in the second equality  ${\cal J}^{\mu}_{ADT}(K)= \nabla_{\nu}{\bf Q}^{\mu\nu}(K)$. Under the condition $\delta K=0$, the generic variation leads to $\delta\Lie_{K}\Psi = \Lie_{K}\delta \Psi$. And thus, the variation of the symplectic current becomes $\delta_{2}\omega^{\mu}(\Psi\,|\, \Lie_{K}\Psi,\delta_{1}\Psi) =\omega^{\mu}(\Psi\,|\, \Lie_{K} \delta_{2}\Psi,\delta_{1}\Psi)$, because of $\Lie_{K}\Psi=0$. By taking into account the second variation of the relation in~Eq.~(\ref{RelsymCurexp}) with the above second variation of the ADT current ${\cal J}^{\mu}_{ADT}$ for the Killing vector $K$, one can see that
\begin{align}   \label{}
&\Big[ \omega^{\mu}(\Psi\,|\,\Lie_{K}\delta_{2}\Psi,\delta_{1}\Psi) + 2\sqrt{-g} {\cal J}^{\mu}_{ADT}(K\,;\, \delta_{2}\Psi, \delta_{1}\Psi)\Big]_{on-shell}  \\
&=  \partial_{\nu}\Big[2\sqrt{-g}{\bf Q}^{\mu\nu}(K\,;\,\delta_{2}\Psi, \delta_{1}\Psi)  + 2(\delta_{2}\sqrt{-g}){\bf Q}^{\mu\nu}(K\,;\,\Psi, \delta_{1}\Psi)   - \sqrt{-g}{\bf A}^{\mu\nu}(\Psi\,|\,\Lie_{K}\delta_{2}\Psi, \delta_{1}\Psi) \Big]_{on-shell} \,,  \nn 
\end{align}
where we have used ${\bf A}^{\mu\nu}(\Lie_{K}\Psi,\delta \Psi)=0$. Under the chosen gauges near the horizon and the asymptotic infinity, it turns out that $\delta\sqrt{-g}|_{B(t)}=0$ at the future horizon and all the terms vanish at infinity because $\delta \Psi$ decays sufficiently fast at infinity. Thus,  one concludes that 
the difference is written eventually as 
\begin{align}   \label{}
&\int_{ \Sigma}dx_{\mu}\Big[ \omega^{\mu}(\Psi\,|\,\Lie_{K}\delta_{2}\Psi,\delta_{1}\Psi) + 2\sqrt{-g} {\cal J}^{\mu}_{ADT}(K\,;\, \delta_{2}\Psi, \delta_{1}\Psi)\Big]_{on-shell}    \nn \\
&=  -\int_{B}dx_{\mu\nu}~ \sqrt{-g}\Big[2{\bf Q}^{\mu\nu}(K\,;\,\delta_{2}\Psi, \delta_{1}\Psi)   - {\bf A}^{\mu\nu}(\Psi\,|\,\Lie_{K}\delta_{2}\Psi, \delta_{1}\Psi) \Big]_{on-shell} \,,
\end{align}
which holds for any Cauchy surface $\Sigma(t)$, not just at $\Sigma(t=0)$.
As a result, one obtains the relation given in Eq.~(\ref{RelHW}). It is interesting to observe that  the above relation   reproduces the same expression given by Eq.~(\ref{GaugeRel}) and Eq.~(\ref{GaugeRel2}) by taking $\delta_{2} = \Lie_{\epsilon}$, which might be just a coincidence not warranted from the construction.
In pure Einstein gravity, one can show, by the explicit computation as done in Appendix B,   that the ${\bf A}^{\mu\nu}$-tensor term does not contribute at the surface $B(t)$ in the chosen coordinates near the horizon as~(\ref{GaussN}) and~(\ref{PertH}). By using the form the perturbed metric in Eq.~(\ref{PertH}) and the fact that $n^{[\mu}\ell^{\nu]}h_{\nu\alpha}\rightarrow 0$ at the future horizon {\it i.e.} at  $r=0$, one can see that the relevant potential term for the horizon Killing vector $K_{H}$  is given by 
\begin{equation} \label{}
{\bf Q}^{\mu\nu}(K_{H}\,;\,\delta_{2}g, \delta_{1}g)\Big|_{B(t)}  =  \frac{1}{2} \delta_{2} g^{\alpha\beta}\delta_{1}g_{\alpha\beta}\nabla^{[\mu}K^{\nu]}_{H} -\frac{1}{2}\delta_{1}g^{\alpha\beta}K_H^{[\mu} \nabla^{\nu]}\delta_{2} g_{\alpha\beta} +K_H^{[\mu} \nabla^{\nu]}(\delta_{2}g^{\alpha\beta}\delta g_{\alpha\beta}) \Big|_{B(t)} \,, \nn
\end{equation}
where we have used that $K_{H}$ is normal to the future horizon  and  $n_{[\mu}\ell_{\nu]}h^{\mu\alpha}\rightarrow0$ as $r\rightarrow 0$. Thus, we obtain
\begin{equation} \label{ReltoHW1}
2n_{[\mu}\ell_{\nu]}{\bf Q}^{\mu\nu}(K_{H}\,;\,\delta g, \delta g)\Big|_{B(t)} = \Big[\kappa\,\delta \mu^{\alpha\beta}\delta \mu_{\alpha\beta} - \frac{3}{2} \delta \mu^{\alpha\beta}\Lie_{K_{H}}\delta \mu_{\alpha\beta}\Big]_{B(t)}\,.
\end{equation}

The above relation may also be written as
\begin{align}   \label{}
&\int_{\partial \Sigma}dx_{\mu\nu}\Big[ \omega^{\mu}(\Psi\,|\,\Lie_{K}\delta_{2}\Psi,\delta_{1}\Psi) + 2\sqrt{-g} {\cal J}^{\mu}_{ADT}(K\,;\, \delta_{2}\Psi, \delta_{1}\Psi)\Big]_{on-shell}    \nn \\
&=  \partial_{\nu}\Big[\delta_{2}\delta_{1}K^{\mu\nu}(K) - 2K^{[\mu}\delta_{2}\Theta^{\nu]}(\delta_{1}\Psi)   -2\sqrt{-g}{\bf Q}^{\mu\nu}(K\,;\,\Psi, \delta_{2}\delta_{1}\Psi)    \Big]_{on-shell}
\end{align}
From the relation in Eq.~(\ref{ADTpot}),  one can see that the last term in  the right-hand side  in the equality cancels  the second order perturbation terms in the proceeding terms. Explicitly, it can be written as
\begin{equation} \label{}
2\sqrt{-g}{\bf Q}^{\mu\nu}(K\,;\,\Psi, \delta_{2}\delta_{1}\Psi)  = \delta_{2}\delta_{1}K^{\mu\nu}(K)\Big|_{\delta_{2}\delta_{1}\Psi} - 2K^{[\mu} \Theta^{\nu]}(\Psi,\delta_{2}\delta_{1}\Psi)\,,
\end{equation}
where the subscript $\delta_{2}\delta_{1}\Psi$ means that we should keep the second order variations.
Schematically, one can write the above relation of the current expression as 
\begin{align}   \label{}
&\Big[ \omega^{\mu}(\Psi\,|\,\Lie_{K}\delta_{2}\Psi,\delta_{1}\Psi) + 2\sqrt{-g} {\cal J}^{\mu}_{ADT}(K\,;\, \delta_{2}\Psi, \delta_{1}\Psi)\Big]_{on-shell}   \\
& =   \partial_{\nu}\Big[\delta_{2}\delta_{1}K^{\mu\nu}(K) - 2K^{[\mu}\delta_{2}\Theta^{\nu]}(\delta_{1}\Psi)   \Big]^{\delta_{2}\delta_{1}\Psi=0}_{on-shell}\,,  \nn 
\end{align}
where $\delta_{2}\delta_{1}\Psi=0$ in the superscript denotes the absence of second order variations in the expressions.

On the bifurcation surface $B$, the contribution comes from the first term $\delta^{2}K^{\mu\nu}$, only. In fact, the essentially same relation has already been obtained in Ref.~\cite{Hollands:2012sf} (see Eq.~(81)  there), though its derivation and interpretation seem to be different. In the end, the difference between ${\cal E}_{HW}$ and ${\cal E}$ is given by 
\begin{equation} \label{}
\CE_{HW}(K\,;\, \delta\Psi,\delta\Psi)- \CE (K\,;\, \delta\Psi,\delta\Psi)  
 =  -\frac{1}{16\pi G} \int_{B}dx_{\mu\nu}~\Big[\delta^{2}K^{\mu\nu}(K)    \Big]^{\delta^{2}\Psi=0}_{on-shell}~,
\end{equation}
which can also be written, through Eq.~(\ref{ReltoHW1}),  at least in Einstein gravity as 
\begin{equation} \label{Diff}
\CE_{HW}(K\,;\, \delta\Psi,\delta\Psi)- \CE (K\,;\, \delta\Psi,\delta\Psi)  
 =  -\frac{1}{8\pi G} \int_{B}dx_{\mu\nu}\, \sqrt{-g}{\bf Q}^{\mu\nu}(K\,;\,\delta \Psi, \delta\Psi)  \Big|_{on-shell}\,.
\end{equation}
By noting that the modified canonical energy differs from the HW canonical energy  only on the bifurcation surface $B$, whenever $\delta_{1}\Psi$ or $\delta_{2}\Psi$ are taken in such a way that $\delta_{1,\,2}Q(K)  = 0$ at infinity,   one can see that our modified canonical energy satisfies the same properties with the HW canonical energy for the perturbation {\it toward stationary black holes}.

\end{document}